\documentclass[12pt,letterpaper]{article}
\usepackage[utf8]{inputenc}
\usepackage{amsmath, amssymb, amsthm, amsfonts, lipsum}
\usepackage[backend=biber, style=apa, annotation=false]{biblatex} %removed for arXiv
\usepackage[english]{babel}
\usepackage{csquotes}
\usepackage{graphicx}
\usepackage{pstricks,rotating}
\usepackage{array}
\usepackage{pdfpages}
\usepackage{setspace, url} % imports the setspace package.
\usepackage{fullpage}
\usepackage{paralist}
\usepackage{verbatim}
\usepackage{lscape,epstopdf}
\usepackage{appendix}
\usepackage{multirow}
\usepackage{xcomment}
\usepackage{subcaption}
\usepackage{hyperref}
\usepackage{ragged2e}
\usepackage{rotating}
\usepackage{enumitem}
\usepackage{titling}

\title{\vspace{-0.5em} Helping People Choose Careers in the Age of AI \vspace{-.5em}}
\author{
	Jennifer L.\ Steele\textsuperscript{1} \and
	Isabella Cruz\textsuperscript{2}
}
\date{July 2026 \vspace{-1em}}

\begin{document}
	
\maketitle

\begin{center}
	\textsuperscript{1}American University, steele@american.edu \\
	\textsuperscript{2}University of Colorado Boulder, isabella.cruz@colorado.edu \\
\end{center}

\begin{abstract}
	\noindent How should people choose careers when artificial intelligence (AI) is rapidly transforming the nature of work? We first compare six recent projections of occupational exposure to task automation with AI, examining their methods and assumptions. We then propose a new empirical model of occupational AI exposure based on 2025 query data from Anthropic and OpenAI. We find marked heterogeneity in model predictions, though models published since 2020 show positive relationships among AI exposure, salaries, and occupational complexity. To reduce uncertainty due to heterogeneous assumptions about task automation potential, we average the projections from five models, including our own. Using these averages, we report on likely tradeoffs between salaries and AI exposure across interest categories, O*NET Job Zones, and job fields. Jobs in healthcare practice show the strongest balance of higher pay with lower AI exposure. Among jobs making high use of Anthropic's Claude, those that use it as a complement rather than a substitute for human work are modestly higher-paying, though whether this pattern holds will depend on usage norms adopted in each field.
\end{abstract}

\vfill
\noindent\rule{0.3\textwidth}{0.4pt} \\
\noindent\textbf{Acknowledgments:} We are grateful for helpful feedback from attendees at the 2024 Fall Research Conference of the Association for Public Policy Analysis and Management, the 2026 Spring Conference of the Association for Education Finance and Policy, and the Spring 2026 Research Brownbag Series in the Department of Public Administration and Policy at American University. Any errors are attributable to the authors.

\newpage

\section{Introduction}
Rapid technological progress in Artificial Intelligence (AI) is quickly redrawing the boundary between work that humans and computers can do. Public reaction runs the gamut from enthusiastic to alarmed, with the former dominating in emerging economies and the latter in advanced ones \parencite{IPSOS2026, Tyrangeil2026, Yang2026}. In a 2025 survey of 47,000 adults across 47 countries, University of Melbourne researchers found that 74\% of Nigerian respondents and 69\% of Chinese respondents believed the benefits of AI outweighed the risks, as compared to only 35\% of U.S. and 30\% of Australian respondents \parencite{Gillespie2025}. The reasons for these differences are a matter of speculation, but people's expectations about the distribution of AI benefits probably play a role. In China, for example, leading Large Language Models (LLMs) like DeepSeek are open-weight models, which reduce costs and increase price transparency for users \parencite{Nagle2025}. Also, there may be a tendency in emerging economies to view AI as infrastructure that enhances the skills of workers rather than as a super-intelligence poised to replace them \parencite{Dreyer2026, Wang2026}. 

Of course, technological innovation has been disruptive throughout history. Tool-building has facilitated human flourishing while advantaging groups that possess contextually optimized tools. The dawn of agriculture 12,000 years ago allowed societies with arable land to produce food surpluses and population growth, giving rise to divisions of labor, social hierarchies, and conflict-laden quests for more land \parencite{Chu2022, Guzman2011}. During the Industrial Revolution two centuries ago, steam engines allowed the movement of goods across long distances, expanding consumer markets but also intensifying competition for small producers. Artisan weavers were replaced by less-skilled power loom workers who toiled under harsh conditions for half the pay of their predecessors \parencite{Acemoglu2024}. More recently, the computing revolution of the late twentieth century automated middle-income manufacturing and routine clerical jobs, while boosting the complementary productivity of non-routine cognitive tasks like computer programming, product design, and data analysis \parencite{Autor2003}. Because lower-skilled jobs could be automated and higher-skilled jobs could not, wages became increasingly polarized by educational attainment \parencite{Autor2013}. Less-educated workers were pushed into the low-paying service sector while highly-educated workers found remunerative opportunities in the knowledge economy \parencite{Acemoglu2011, Goldin2008}.

Today, innovations in AI presage a new disruption. As of 2025, 21\% of U.S. workers said they used AI tools such as LLMs for at least some of their work \parencite{Lin2025}. A spring 2025 Anthropic study that linked Claude usage queries to occupational codes found that 36\% of U.S. occupations were using Claude for at least a quarter of their tasks \parencite{Handa2025}. In May 2026, the job board Indeed.com reported that 5.7\% of US job postings included language related to AI (e.g., "machine learning," "large language models,"), representing more than a threefold increase over the preceding three years \parencite{Indeed2026}. 

Still, the current impact of AI on employment remains a matter of empirical debate. Using job posting data from 2010 through 2018, \textcite{Acemoglu2022} reported increased AI-related job openings and concurrent declines in other job postings among firms with high AI exposure as defined by three distinct AI exposure-projection models we consider in this paper \parencite{Brynjolfsson2018, Felten2021, Webb2020}. \textcite{Demirci2025} found that between 2023 and 2025, freelance job postings in writing and programming dropped by approximately 21\% consistent with early substitution effects in writing and coding, skills in which LLMs excel. \textcite{Brynjolfsson2025} used U.S. payroll data and AI exposure estimates from both OpenAI and Anthropic \parencite{Eloundou2024, Handa2025} to detect employment declines of up to 13\% for early-to-mid-career workers in occupations with the highest exposure to tasks that can be fully automated (not just assisted) by AI \parencite{Brynjolfsson2025}. Their findings suggested that firms were replacing entry-level knowledge workers with AI, as many observers had predicted \parencite{Beane2024}. 

However, newer papers have questioned whether young workers' unemployment is attributable mainly to generative AI or to the rise of remote work, which has been shown to increase the logistical cost of training junior workers \parencite{Emanuel2026}, and which slightly predated the rise of generative AI \parencite{Lambert2026}. Using data on job postings and new hires in the U.S., UK, Canada, and Australia from 2017-2025 \textcite{Lambert2026} show that remote work arrangements offer a better explanation of falling entry-level employment, though jobs that can be done remotely and those that can be done by generative AI show a correlation of 0.7 in their study. A study by \textcite{Schubert2025} using  U.S. job postings data from Lightcast suggests the correlation is not incidental. He finds that a 10 percentage- point increase in firms' shifts to remote work during the pandemic predicts a subsequent 0.4 percentage-point increase in AI mentions in job postings, arguably because the atomization and formalization of in-person tacit knowledge that facilitates remote work also streamlines the delegation of tasks to AI. 

A recent paper by the financial services company Ramp and the workforce analytics company Revelio Labs examined AI investments and employee headcounts across nearly 22,000 firms \parencite{Kharazian2026}. Among firms that began spending at least \$100 per month on AI software or tools, those in the highest-spending tercile (averaging \$33 per month per employee) showed 10.2\% growth in headcounts in the two years after AI spending commenced, driven mainly by entry-level roles, including roles in sales and customer service. Lower-spending terciles, who spent about \$3 per month per employee on AI tools, did not show a statistically significant change in headcounts. Though they attempted to control for exogenous company growth rates by limiting the control group to firms that eventually adopt AI, finding such large headcount growth for fairly low levels of AI spending -raises questions as to whether the effects can be causally attributed to AI investments.

Even if early effects of AI tools on the labor market remain ambiguous, the growing ease with which advanced coding, writing, analytic, and design tasks can be automated raises profound questions about the future of knowledge work. Though decisions about the future are uncertain by definition, the rapid growth in AI capacity has added new risks to education and career choices. Our analysis in this paper takes a risk-diversification approach in which we compare the properties of six prominent sets of AI exposure projections and develop our own set of projections based on empirical usage data from Anthropic and OpenAI. We then synthesize these projections, including our own, to provide a cleaner signal for the career-investment decisions of students, career-changers, and those advising them, including college and career counselors. We show where the projections differ and converge in terms of the occupational interest categories, salaries, and education levels, and we identify the occupational fields that pair above-median salaries with higher and lower projected exposure to AI. Readers who wish to explore the data for particular jobs and AI projection models can use our interactive data tools, available at \url{https://tinyurl.com/SteeleCruzCareersAI}.

\subsection{A Framework for Technology and Human Complementarity}
Cautious optimism about the effects of AI on workers have been driven by Jevons paradox: the observation that making a resource more efficient can paradoxically increase its total consumption if demand is sensitive to price \parencite{Jevons1865}. The paradox is easy to understand in the case of physical resources like fuel, which was its original application. But it can apply to labor demand during automation if consumers' demand for cheaper tasks increases with falling prices \textit{and} if human skills are necessary complements to the tasks being automated \textcite{Bessen2019, Autor2024}. Both factors are likely to vary by industry. Focusing on the complementarity of AI and human skills, \textcite{Brynjolfsson2022} argues that humans should prioritize digital applications that augment rather than supplant human advantages, and that ``[t]he distributive effects of AI depend on whether it is primarily used to augment human labor or automate and replace it.''

To formalize the complementarity of humans and technology, consider an occupation that produces reports using a technology input $A$ (e.g., software) and a human skill input $H$ (e.g., analyst hours). Output $Y$ is a function of $A$ and $H$, whose relative importance weights are specified by $\alpha \in (0,1)$ and its complement, $(1-\alpha)$. Elasticity of substitution, meaning the ease with which technology and human labor are substituted for one another, is a function of substitution parameter $\rho < 1$, which implies a constant elasticity of substitution $\sigma = \frac{1}{1-\rho}$. The production function is specified in equation~\eqref{eq1}:

\begin{equation}
Y = [(\alpha)A^{\rho} + (1-\alpha)H^{\rho}]^{1/\rho}
\label{eq1}
\end{equation}

Plausible values of $\rho$ and $\sigma$ depend on how easily labor inputs can be substituted for one another within a given occupation. When $\rho$ is positive and less than 1 ($0 < \rho < 1$), $\sigma > 1$, meaning that technology can be easily substituted for human skill and vice versa. This happens, for instance, when real-time language translation software substitutes for human translators at public events, or when chatbots rather than copywriters produce marketing copy for websites. When $\rho = 0$, $\sigma = 1$, meaning substitution is proportional but involves friction, as when AI-generated newsletters require human fact-checking and editing. Finally, when $\rho < 0$, this indicates that $\sigma < 1$, meaning that technology and human skills are complementary, as when an analyst needs software to produce a data-analytic report, or an anesthesiologist needs equipment to monitor patients during surgery. As $\rho$ approaches negative infinity, $\sigma$ approaches 0 ($\lim_{\rho \to -\infty} \sigma = 0$), meaning that technology and human skill become nearly perfect complements and cannot replace one another. 

As a stylized example, let $\alpha = 0.5$, meaning that technology and human labor are equally weighted in importance, and $\rho = -1$, indicating low elasticity of substitution, so that:
\begin{equation}
Y = [0.5A^{-1} + 0.5H^{-1}]^{-1}
\label{eq2}
\end{equation}
This specification implies that $\sigma = 1/2$, capturing strong complementarity between technology and human skill. When this is true, output $Y$ declines as $A$ or $H$ becomes small.

Suppose a firm initially uses $A = 10$ units of technological inputs and $H = 10$ units of human input. Output is then:
\begin{equation}
Y_0 = [0.5 \cdot 10^{-1} + 0.5 \cdot 10^{-1}]^{-1} \\
= [0.05 + 0.05]^{-1} \\
= 0.1^{-1} = 10 
\label{eq3}
\end{equation}

At this point, the occupation produces $Y_0 = 10$ reports. But suppose an AI upgrade doubles the efficiency of the technological component. We can represent this by a factor $\theta = 2$, with effective technology $\tilde{A} = (\theta) A$. If the firm keeps the same raw technological effort,
$A=10$, then $\tilde{A} = 20$ and $H$ remains $H=10$. Plugging these inputs into the production function yields:
\begin{equation}
Y_1 = [0.5 \cdot 20^{-1} + 0.5 \cdot 10^{-1}]^{-1}  \\
= [ 0.025 + 0.05]^{-1} \\
= 0.075^{-1} \approx 13.33
\label{eq4}
\end{equation}

Improved technological efficiency thus raises output from $Y_0 = 10$ to $Y_1 \approx 13.33$ for the given inputs, indicating a productivity increase of about 33\%. 

However, if the elasticity of substitution were higher, making it easier to substitute technology for human skill, productivity with ($\tilde{A}$) would rise further because the gains would not depend greatly on the human component. For instance, if $\rho = 0.8$ and $\sigma = 5$, indicating high elasticity of substitution, then:
\begin{equation}
Y_2 = [0.5 \cdot 20^{0.8} + 0.5 \cdot 10^{0.8}]^{1.25} \\
= [5.493 + 3.155]^{1.25} \\
= 8.648^{1.25} \approx 14.83
\label{eq5}
\end{equation}
In other words, output would rise from $Y_0 = 10$ to $Y_2 \approx 14.83$ due to less dependency of the technological gains on human skill, even before any reallocation of resources. This occurs because the increase in technology capacity is captured in output with minimal dependence on complementary human capacity \parencite{Vivanco2018}. Technological tasks can be delegated with less coordination and thus have greater likelihood of displacing complementary human workers. The point is that Jevons paradox predicts expanded occupational demand when the demand for the bundle of tasks in the occupation is price-elastic \textit{and} the ease of substituting technology for labor is relatively \textit{inelastic} \parencite{Acemoglu2011, Hernnas2023}. 

As \textcite{Agrawal2024} have pointed out, reduced demand for expert skills may open opportunities for workers in other parts of the skill distribution. They argue that AI may enhance economic equality by reducing entry barriers to high-skill tasks for workers with lower skills, as in the case of GPS-assisted Uber driving or Claude Code assisted vibe-coding. "One worker's automation," they write, "is another's augmentation," suggesting that what one worker might offload to AI as drudgery, such as drafting or proofreading emails, another might gain the capacity to do because of AI assistance. \textcite{Autor2025} reinforce this argument, showing that the effect of task automation on employment and wages depends not on how many tasks are automated, but on whether the automated tasks are expert or inexpert tasks within the occupation. They show that automating expert tasks expands employment opportunities to more workers but reduces wages, whereas automating inexpert tasks does the reverse. \textcite{Klein2026} reinforce this finding with a study of job postings across 39 countries, in which the launch of ChatGPT predicted slightly higher wages in the next 2.5 years for jobs in which lower-skill tasks were AI-exposed.  

\begin{figure}[htbp]
	\centering
	\begin{minipage}{0.49\textwidth}
		\centering
		\caption{AI optimism versus senior salaries across 37 countries}
		\label{benexsal}
		\includegraphics[width=\textwidth]{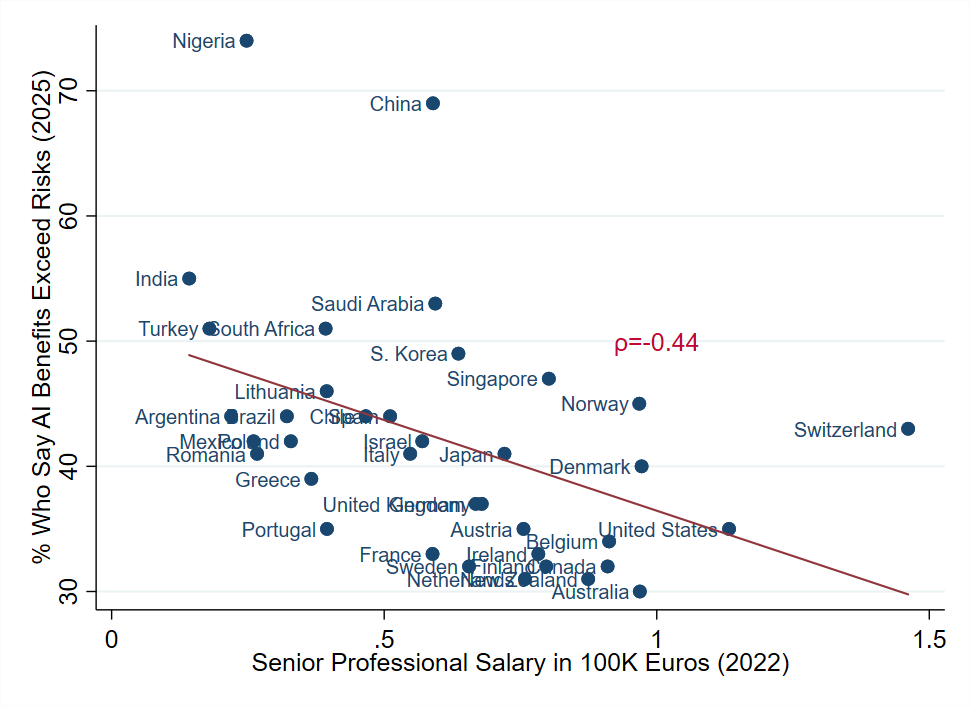}
	\end{minipage}\hfill
	\begin{minipage}{0.49\textwidth}
		\centering
		\caption{AI optimism versus professional salary-to-minimum wage ratios}
		\label{benexrat}
		\includegraphics[width=\textwidth]{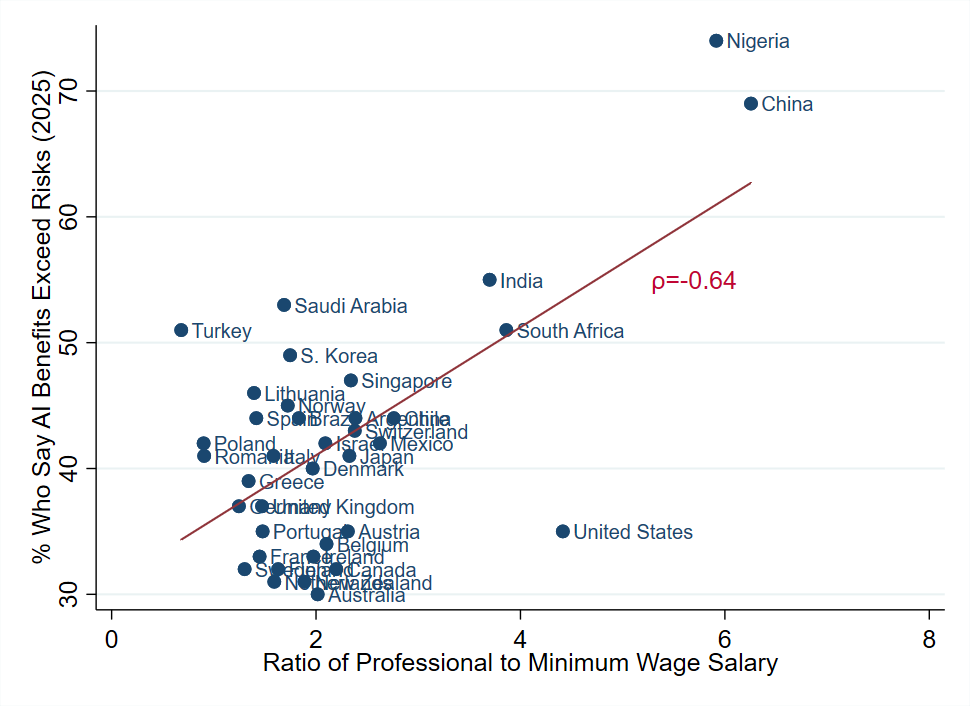}
	\end{minipage}
\end{figure}

Belief in the equalizing power of AI tools may underpin aforementioned differences in AI optimism between developed and developing economies. We show in Figures \ref{benexsal} and \ref{benexrat} that the average 2022 salaries of senior professionals across 37 countries \parencite{N262022} is negatively related to the country's AI optimism in 2025 ($\rho=-0.44$), defined as the percent of residents surveyed by \textcite{Gillespie2025} who believe the benefits of AI exceed the risks. In contrast, a country's salary inequality, measured here by the ratio of junior-level professional salaries to the minimum wage in 2022 \textcite{N262022}, is positively related to AI optimism, with a correlation ($\rho$) of 0.64. Both relationships to AI optimism are statistically significant even when controlling for the fraction of people in the country who hold postsecondary degrees, as shown in Appendix Table \ref{tabregress}. 

Consistent with this optimistic view, recent experimental evidence suggests that lower-skilled workers may have the most to gain from access to AI tools. \textcite{Noy2023} randomly assigned 453 marketers and grant writers to ChatGPT versus business-as-usual, finding that ChatGPT access increased writing speed, quality, and enjoyment, with larger benefits for less-skilled writers. In a 5,000-person call-center experiment studied by \textcite{Brynjolfsson2023}, access to AI-generated customer support scripts yielded a 14\% increase in call resolutions per hour, again with larger gains among lower-skilled workers. These data contradict the argument that entry-level work is the most vulnerable to AI displacement. On the other hand, the equalizing potential of AI depends on workers remaining integral to the efficiencies it extracts.

The preceding discussion illustrates why an occupation's exposure to AI does not necessitate job displacement and why the displacement of particular skills by AI may expand opportunity to a broader array of workers. Such expansion could have benefits across the skill distribution, if, for example, medical schedulers can book patient appointments with greater ease and top physicians can improve diagnostic speed and accuracy, serving more patients while maintaining or improving quality of service. Even so, AI exposure may change the composition of tasks within jobs and force workers to adapt the way they work. Given the possibility of job change as well as displacement, there is value in providing greater clarity to students and career changers about likely occupational exposure to AI.

\section{Contributions of the Current Paper}
Guided by the aim of providing clarity to career seekers, our paper synthesizes six extant projections of how AI will affect individual careers and fields of interest. We describe the distinctive approaches of each prior model, including their assumptions and methods, and we document heterogeneity in their predictions. Building on insights from their approaches, we create an empirically grounded exposure measure using task-level data from Anthropic spanning August 4-11, 2025 \textcite{Appel2025}, and Generalized Work Activity (GWA)-level data from OpenAI spanning  May through July 2025 \textcite{Chatterji2025}. Our measure is correlated with Anthropic's exposure projection-model by \textcite{Massenkoff2026}, but our approach departs from that of \textcite{Massenkoff2026} in that we also incorporate OpenAI usage data from \textcite{Chatterji2025}, and we define plausible AI usage boundaries for tasks as a function of known usage patterns. We highlight how Claude and ChatGPT usage patterns differ and why it is therefore useful to incorporate data from both foundational models.

Comparing our own model to that of \textcite{Massenkoff2026} and five other prominent AI-exposure models, we document considerable heterogeneity in model projections. This heterogeneity complicates advising for career seekers. To reduce uncertainty in predictions due to heterogeneous assumptions about task and activity automation exposure, we calculate a cross-model average exposure score based on our own model and four extant models that use different but complementary approaches. In addition, we examine whether task automation in Anthropic data shows a different relationship to salaries than task augmentation, demonstrating that higher-paying jobs have a slightly higher propensity toward augmentation than full automation. 

Given that jobs' AI exposure is positively correlated with salaries in most projections, we illustrate how occupations, fields of interest, and job sectors vary in terms of two important dimensions: median salaries and projected AI exposure, highlighting the career fields and sectors that balance higher salary levels with lower projected AI exposure. Because our cross-model estimates incorporate the assumptions of five independent projection models, they are intended to inform career decision-making by career seekers, college and career counselors, and educational institutions in the era of generative AI.

\section{Recent Models Predicting Jobs' AI Exposure}
Table \ref{tabmodel} summarize the estimates we compare in this paper, from newest to oldest. First shown is our own approach, which, as described in subsequent sections, is based on observed data from Anthropic and OpenAI usage in 2025. Next listed is the \textcite{Massenkoff2026} measure of observed occupational exposure, which focused on work-related Anthropic Claude queries from August and November 2025 as linked to 17,998 O*NET tasks. They calculated a task's exposure to AI as ranging from 0 to 1, where 0 is a task that accounts for less than 0.0025\% of Claude traffic (<100 queries) \textit{or} has a $\beta$ value of less than 0.5 (about 44\% of tasks) in the theoretical exposure model by \textcite{Eloundou2024}, which is described below. An exposure score of 0.5 by \textcite{Massenkoff2026} means that tasks done by Claude that were deemed feasible by \textcite{Eloundou2024} were entirely augmented (that is, human-controlled) rather than automated. The total \textcite{Massenkoff2026} task-exposure value (technically, $[0.5+[0.5\cdot(automated\_share)]$) can be interpreted as the fraction of an LLM-feasible task that Claude was handling in late 2025. The authors aggregated these task-level estimates to the occupation level after weighting them by an estimate of occupational time spend on that task, though they report that the estimates were insensitive to this weighting. Their resulting occupation-level estimates ranged from 0 to 0.73, with a mean of 0.08 and a standard deviation of 0.12.

\begin{table}[htbp]
	\centering
	\small
	\caption{Seven AI-exposure models examined in this paper}
	\label{tabmodel}
	\begin{tabular}{ m{0.27\textwidth}  m{0.27\textwidth}  m{0.41\textwidth} }
		\hline
		\textbf{Article} & \textbf{How} & \textbf{Measure} \\
		\hline
		\textcite{Steele2026} (this paper) & Anthropic \& OpenAI queries & \% of job tasks that can be feasibly automated \\
		\hline
		\textcite{Massenkoff2026} & Anthropic queries \& Eloundou feasibility & Share of LLM-feasible tasks fully automated by Claude \\
		\hline
		\textcite{Eloundou2024} & GPT task ratings (human-validated) & Share of job tasks LLM can fully automate  \\
		\hline
		\textcite{Felten2021} & Crowd-sourced job ability ratings & Job's standardized suitability for LLM automation \\
		\hline
		\textcite{Webb2020} & Text-mining of AI patent filings & \% of AI-patent noun-verb pairs in the job's tasks \\
		\hline
		\textcite{Brynjolfsson2017} & Crowd-sourced tasks with rubric  & Job's suitability for machine learning (SML) \\
		\hline
		\textcite{Frey2017} & Human ratings of AI bottlenecks & Probability job can be fully automated \\
		\hline
	\end{tabular}
\end{table}
		
The next model we consider is from \textcite{Eloundou2024}---the partial basis of estimates from \textcite{Massenkoff2026}---who use human raters and GPT 4.0 to evaluate 19,265 O*NET occupational tasks according to whether an LLM in mid-2023 could complete them as well an average worker in half the time. Their paper constructed four measures of LLM exposure: $\alpha$ represents LLM capability alone, $\gamma$ represents tasks LLMs could do by harnessing additional software (not robots), and their preferred measure, $\beta$, reflects the LLM-only score plus half the software-supplemented LLM capability. For their fourth measure, \textit{automation}, they used OpenAI's GPT 4.0 to code occupational tasks for the degree to which LLMs can fully automate them, ranging from 0 for not at all to 1 for tasks that LLMs can do completely, incremented in units of 0.25. At the task level, \textit{automation} and $\beta$ had a correlation of about 0.94. Their job-level automation score, which aggregates task automation scores to the job level for 923 O*NET occupations, can be interpreted as the share of a job that can be fully automated by an LLM. Given this ease of interpretation, we use their job-level automation scores in our cross-model comparisons in this paper. (Note that \textcite{Eloundou2024} weighted tasks equally but, like \textcite{Massenkoff2026}, showed that their results were not sensitive to task weighting.) Across occupations, the \textcite{Eloundou2024} occupational automation score ranged from 0 to 0.84, with a mean of 0.32 and a standard deviation of 0.19.

The next AI exposure model we compare is by \textcite{Felten2021}, who defined exposure based on 10 rapidly advancing AI applications such as reading comprehension; speech, image, and music recognition; translation; and abstract strategy games. Surveying 2000 U.S.-based Mechanical Turk respondents (200 per AI application), they considered whether each of these high-growth applications could be used for each of 52 O*NET occupational abilities. They then aggregated respondents' exposure ratings of abilities to the job level, weighted by the abilities' importance and prevalence within each job. Their AI occupational exposure (AIOE) scores for 774 occupations were then standardized to mean 0 and standard deviation 1. The observed range among occupations was from -2.67 to 1.53.

Taking a different approach, \textcite{Webb2020} defined AI in terms of two types of machine learning: supervised learning, in which models are pre-trained on labeled data (LLM pre-training being an example) and reinforcement learning, in which models update their problem-solving strategies based on past successes and failures. Using these definitions, he extracted verb-object pairs from AI-related patents titles in the Google Patents Public Data. Employing a semantic hierarchy for the objects (nouns), he matched these verb-object pairs to conceptually similar pairs he extracted from about 18,000 O*NET task descriptions across 964 occupations. He aggregated the task exposure scores to occupations using the average of each task's frequency, importance, and relevance weights within each occupation. His resulting measure for each job, the AI score, represents the average percent of AI-patent extracted pairs that appear in the tasks of that job, with a range from 0 to 1.49, a mean of 0.42, and a standard deviation of 0.29. 

\textcite{Brynjolfsson2018} used the crowd-sourcing platform Crowdflower to survey workers about the suitability for machine learning of 2069 O*NET Detailed Work Activities (DWAs) linked to 923 occupations. Their AI exposure measure, called suitability for machine learning (SML), was defined as a function of 23 equally weighted factors including machine-readability of outputs and tasks, complexity and abstraction of tasks, routineness and rule-based nature of tasks, data availability for tasks, and speed and stakes of outputs, with lower-stakes tasks deemed more suitable for machines than high-stakes tasks. Their model linked DWAs to 18,939 job tasks, and then averaged task scores to the job level, weighted by the importance of each task. Because task importance scores were available for only 659 of the 923 occupations, this provided exposure ratings for 659 jobs. In our replication of their analysis, we find that weighting by task importance has virtually no effect on the distribution of job-level exposure (the correlation between the two is 0.986), so we weight the task scores equally in aggregating to the occupation level. This allows us to calculate the \textcite{Brynjolfsson2018} measure for 923 occupations. It has an observed range across occupations of 2.83 to 3.91 on a scale in which 1 means not suitable for machine learning and 5 means highly suitable. It has a mean of 3.46 and a standard deviation of 0.11.

In the earliest model we consider, \textcite{Frey2017} hand-coded 70 jobs dichotomously for susceptibility to full automation given their O*NET task descriptions. Using nine O*NET variables regarding the social skills (perceptiveness, negotiation, persuasion, caring for others), creativity (arts and originality), and perception and manipulation (finger and manual dexterity and awkward positioning) involved in each job, they predicted the probability that the job could be fully automated using an exponential quadratic machine-learning model. They then applied the model parameters to all 702 six-digit O*NET occupations in the dataset. Their resulting occupational exposure ratings, ranging from 0.003 to 0.99, with a mean of 0.5 and a standard deviation of 0.38, can be interpreted as the probability that a job can be fully automated given its exposure to those nine hard-to-automate O*NET skills.

\subsection{Substantive differences among models}
It is important to note the historical progression in these measures. The model by \textcite{Frey2017} draws on assumptions similar to those of older automation waves, including \textcite{Autor2013z} and \textcite{Autor2003}, reflecting the effects of routinized manual and office labor on lower-skilled jobs and predating the creative capacity of LLMs. The model by \textcite{Brynjolfsson2018} also predates LLMs but is based on a 23-point rubric about machine learning capabilities. However, contrary to more-recent models, it assumes that high-stakes tasks are not highly suitable for machine learning. The model by\textcite{Webb2020}, in emphasizing tasks that align with AI patents, places higher emphasis on hardware-enabled innovations, which developers are likely to patent, and less emphasis on tasks that can be automated by foundational LLMs such as ChatGPT, Claude, and Gemini. The models by \textcite{Felten2021} and \textcite{Eloundou2024} use crowd-sourcing and GPT-scoring, respectively, to identify tasks suitable for LLMs and image-generation. \textcite{Massenkoff2026} employs tasks' actual Claude usage, combined with their LLM-suitability as defined by \textcite{Eloundou2024}, to quantify jobs AI exposure. Because the \textcite{Massenkoff2026} exposure measure is dependent on the \textcite{Eloundou2024} measure in addition to reflecting tasks' exposure to Claude queries, we propose an alternative Claude-and-ChatGPT-based exposure measure that is independent of the \textcite{Eloundou2024} measure and instead bases likely exposure on the \textit{intensity} of tasks' exposure to Claude and ChatGPT, as we explain below.

\begin{table}[htbp]
	\centering
	\caption{Raw distributions of model estimates}
	\label{tabcomp}
	\begin{tabular}{lccccc}
		\hline
		Variable & Obs & Mean & Std.\ Dev. & Min & Max \\
		\hline
		Steele (2026)  	  & 872 & 23.6 & 6.18 & 15.8 & 49.8   \\
		Massenkoff (2026) & 872 & 0.08 & 0.12 & 0 & 0.73 \\
		Eloundou (2024)   & 872 & 0.32 & 0.19 & 0  & 0.84   \\
		Felten (2021)     & 759 & 0.04 & 1.0  & $-$2.67 & 1.53 \\
		Webb (2020) 	  & 719 & 0.42 & 0.29 & 0 & 1.49  \\
		Brynjolfsson (2018) & 872 & 3.46 & 0.11 & 2.83 & 3.91  \\
		Frey (2017)		 & 689 & 0.50 & 0.38 & 0.003 & 0.99 \\
		\hline
	\end{tabular}
\end{table}

Table \ref{tabcomp} summarizes the raw distribution of automation exposure estimates from each of the models just described, in addition to our own model, which we explain below. The maximum number of O*NET jobs in our analysis is 872 out of a paper-specific maximum of 923, due to slightly different Standard Occupational Code versions used across the papers. Projections from \textcite{Steele2026}, \textcite{Massenkoff2026}, and \textcite{Eloundou2024} can be interpreted as the percentage or fraction of the job tasks exposed to AI-enabled automation, and \textcite{Frey2017} can similarly be interpreted as the probability that a given job can be fully automated with AI. Frey's is the only model that classifies jobs as unitary entities rather than baskets of tasks, activities, or abilities. 

In the aforementioned study by \textcite{Acemoglu2022} of U.S. job postings between 2010 and 2018, the AI-exposure predictions of \textcite{Felten2021} were somewhat more effective than those of \textcite{Webb2020} and \textcite{Brynjolfsson2018} at predicting growth in firms' posting of AI-related jobs, but the study data preceded the November 2022 release of ChatGPT 3.5 so may not fully anticipate the rise of LLMs. 

In the ensuing discussion, we describe the construction of our own empirically based AI exposure model using Anthropic and OpenAI data. We then consider how all of the model projections correlate with one another before proposing a cross-model composite measure.
	
\section{Our Query-Based Exposure Measure}
To construct our own measure of occupational exposure, we begin with task-level usage data published by \textcite{Appel2025} in September of 2025 based on Anthropic's classification of global queries sent to Claude AI tools during August 4-11, 2025.\footnote{Data are from the third Anthropic Economic Index report (Appel, 2025), with data available at \href{https://huggingface.co/datasets/Anthropic/EconomicIndex}.} The Anthropic data include 1,909,132 global queries made to Claude, in which 50.5\% of queries were sent to Claude Free or Pro and 49.5\% were sent to Claude's Application Programming Interface (API), which is used mainly as part of business workflows. We combine the Claude Free/Pro and API usage data, both of which assign each query to one of 17,659 O*NET tasks based on the content of the query. Each query is also sub-classified by Claude as conforming to one of five usage patterns: (1) \textit{validation}: asking Claude to critique or edit something the user has produced; (2) \textit{learning}: asking Claude to explain or teach the user something; (3) \textit{iteration}: engaging in a conversation with Claude around refinement of an idea or work product; (4) \textit{feedback loop}: assigning a task to Claude that it performs with only small tweaks or input from the user; and (5) \textit{directive}: assigning a task to Claude that it completes on its own. The Anthropic Economic Index characterizes usage patterns 1-3 (validation, learning, and iteration) as \textit{augmentation} of work, since the user controls how the task is performed by Claude. Anthropic characterizes patterns 4-5 as \textit{automation} of work, in which agency over task execution is delegated largely or entirely to Claude. The September 2025 data release shows that about 40\% of Claude Free/Pro queries and 70\% of Claude API queries qualified as task automation rather than augmentation. The former figure is similar to the 43\% of Claude Free/Pro tasks that were automated (as opposed to augmented) in January 2025 \parencite{Handa2025}.

\subsection{Extrapolating from Task Usage to Exposure}
In constructing an AI-exposure measure based on Anthropic usage in August 2025, we follow \textcite{Appel2025} in calculating the number of tasks that are automated by adding the number of directive and feedback queries across both Claude Free/Pro and API queries for each of the 17,659 tasks. We calculate the number of tasks that are augmented by adding the number of Claude Free/Pro and API queries classified as validation, learning, iteration, as well as queries whose purposes are deemed unclassified. The total number of queries per task is thus the sum of the automated and augmented queries. 

Because each task may appear in multiple occupations, but each occupational-task pairing is unique, we aggregate the total number of queries to the occupational level and divide that occupational total by the 1,909,132 Claude Free/Pro and API queries assigned to tasks in the August 4-11, 2025 dataset. Multiplying by 100 yields the percentage of queries assigned to each of 872 O*NET occupations. This occupational percentage has a range from 0 to 14\%, with a mean of 0.14\% and a median of less than 0.01\%. Clearly, the fraction of Claude queries linked to a given job is not a reasonable proxy for the fraction of tasks in the job that AI can feasibly perform given current capacity and the rate of model improvement.

To construct a measure of occupational AI exposure from these data, we must make assumptions about the extent to which tasks can be undertaken by current and near-future extensions of generative AI, including models enhanced with retrieval-based contextual search (e.g., Retrieval-Augmented Generation), advanced multi-step logical reasoning, and agentic capabilities such as task decomposition, iterative planning, and invocation of tools such as code, databases, and APIs. To do so, we first divide task-level total queries into ventiles. Because the distribution of queries-per-task shows a strong positive skew, we classify tasks into four high-use ventiles (17 through 20) and the lowest-use ventile (1), as shown in the first three columns of Table \ref{tabvent}. The lowest-use ventile comprises the 85\% of tasks that have no associated Claude queries. 

\begin{table}
	\centering
	\caption{Claude task query ventiles and projected AI exposure}
	\label{tabvent}
		\begin{tabular}{lcccc}
			\hline    
			Ventile &  n Tasks  & Mean Queries & Projected Exposure & 98\%ile $Aug-Aut$  \\
			\hline
			20 & 643 & 1,829 & 80\% & 40\% \\
			19 & 828 & 103 & 70\% & 40\% \\
			18 & 739 & 29 & 60\% & - \\
			17 & 220 & 16 & 50\% & -\\
			 1 & 15,229 & 0 & 5\% & - \\
			\hline 
		\end{tabular}
\end{table}

We then classify the fraction of a task that could be feasibly automated based on Claude usage as 80\% for ventile 20 queries, 70\% for ventile 19 queries, 60\% for ventile 18 queries, and 50\% for ventile 17 queries, as shown in Column 4 of Table \ref{tabvent}. Based on the assumption that generative AI is well-adapted for the tasks for which Claude is most-often invoked, this coding scheme assumes that tasks making up the top 5\% of Claude queries could be 80\% completed by enhanced AI tools; that those making up the second-highest ventile could be 70\% completed by enhanced AI tools; and that 60\% and 50\% of tasks in the third and fourth-highest ventiles could be accomplished with AI. For the 85\% of tasks with no associated Claude queries, we set the exposure value at 5\% rather than 0\% under the assumption that generative AI could provide useful tips and information that could conceivably absorb up to 5\% of the effort of almost any task. Even highly physical and idiosyncratic tasks like tree-trimming or kitchen sink repair, for instance, might be enhanced by an AI-generated synthesis of best practices. 

On the other hand, the \textit{way} that users employ Claude may also provide clues as to the future AI exposure of particular tasks. Bearing this in mind, we downgrade the AI-exposure potential of tasks that are disproportionately augmented by Claude rather than fully automated, as shown in Column 5 of Table \ref{tabvent}. Specifically, for all tasks in the 98th percentile or higher in terms of the difference between number of augmented versus automated queries (all 223 of which fall in the highest two ventiles of Claude usage), we downgrade potential AI exposure from 70-80\% to 40\%. We do this based on the assumption that these highly exposed tasks would be difficult to automate by more than 40\% due to intrinsic features of the tasks, which in the data deal mainly with teaching, counseling, and scholarly writing. We choose the 98th percentile of the  $Augmentation \- Automation$ query differential because the differences are negative or zero below the 91st percentile and are trivially small just above it. It is only at the 98th and 99th percentiles that they differ from the median of 0 by 0.09 and 0.27 standard deviations, respectively. Substantively, this is because the number of augmentation and automation queries is quite similar for most tasks, making the augmentation versus automation distinction useful mainly for the small percentage of tasks that strongly favor one or the other.

\subsection{Activity Use of Claude versus ChatGPT}
This task-aggregation method works for Anthropic data, but Claude is not the only major foundational AI model, and there is evidence that people use different foundational models for different purposes \parencite{Chatterji2025, Handa2025}.

\begin{figure}[htbp]
	\centering
	\caption{Percent of queries by Generalized Work Activity (GWA)}
	\label{ele_clagpt2}
	\includegraphics[width=0.98\textwidth]{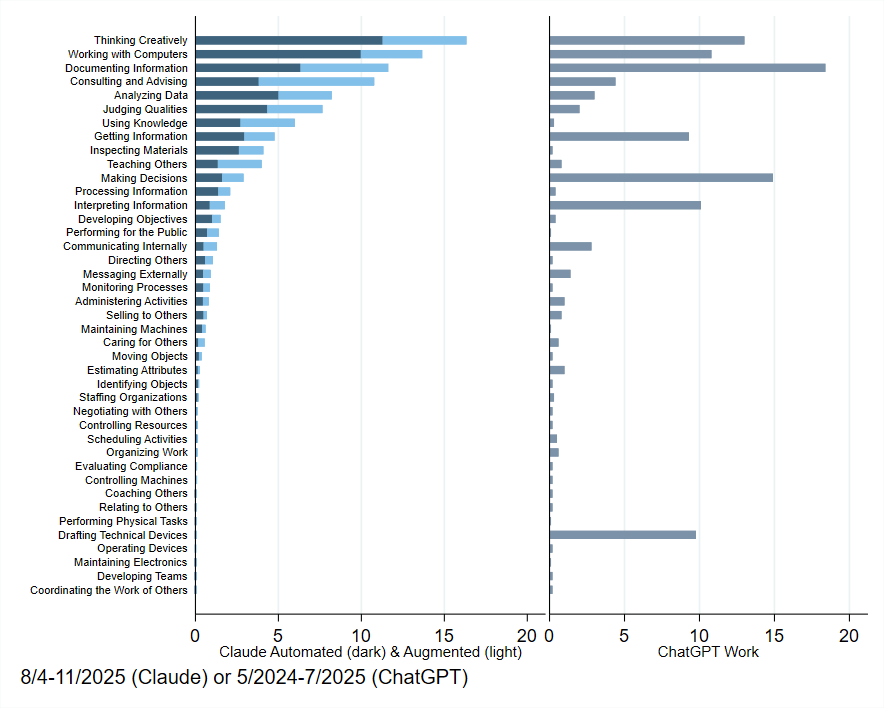}
\end{figure}

We can see these differences clearly in Figure \ref{ele_clagpt2}. In the left panel, we have aggregated Claude Free/Pro and Claude API task queries to the level of Generalized Work Activities, of which there are 41 in O*NET. For tasks that appear in more than one GWA, we divide their queries equally across GWAs to which they are linked. Each bar in Figure \ref{ele_clagpt2} represents the percent of all queries classifiable into each GWA. Queries where Claude is used for validation, learning, iteration, or an unclassified purpose are shown in light blue as "augmented." Those in which Claude executes tasks autonomously or nearly autonomously are shown in dark blue as "automated." 

In the right panel, we show the percent of ChatGPT queries by GWA as reported by \textcite{Chatterji2025} in their ChatGPT usage analysis, but sorted by Claude use prevalence. We focus here on ChatGPT usage in the workplace, which \textcite{Chatterji2025} provided separately in their analysis, and which was not available for Claude usage in the September 2025 release. In the four GWAs for which ChatGPT usage was not reported by \textcite{Chatterji2025} due to the O*NET activity-to-task crosswalk they employed, we impute ChatGPT usage from similar GWAs. For instance, drafting and specifying technical equipment is imputed as the mean of mechanical equipment repair, documenting and recording information, and interacting with computers. Repairing electrical equipment is imputed from repairing mechanical equipment. Coordinating others and building teams are both imputed from guiding, directing, and motivating subordinates. 

In the two panels of Figure \ref{ele_clagpt2}, we observe key similarities between Claude and ChatGPT usage. For instance, thinking creatively, working with computers, and documenting information rank highly for both tools. ChatGPT, however, also ranks very highly for making decisions, interpreting information, and getting information, whereas Claude also ranks highly for interpersonal tasks such as consulting and advising, as well as teaching others. However, as noted, these interpersonal activities---consulting, advising, and teaching---are among the few GWAs for which a larger share of the queries are augmented rather than automated. This is important because the heavy use of Claude for teaching and advising challenges common assumptions, including those in \textcite{Brynjolfsson2018} and \textcite{Frey2017}, about the intrinsic interpersonal advantages of humans over machines. If users are primarily employing Claude to improve the efficiency of their consultations or teaching so they can preserve more time for the interpersonal components, this would be consistent with a human-centered approach to AI \parencite{Brynjolfsson2022}. It is also noteworthy how rarely Claude and ChatGPT were used for other types of interpersonal tasks, including directing others, caring for others, negotiating with others, relating to others, developing teams, or coordinating the work of others. These patterns are consistent with assumptions in \textcite{Frey2017} and \textcite{Brynjolfsson2022} that humans have a natural advantage over machines in understanding and relating to their fellow human beings.

\subsection{Aggregating Claude and ChatGPT Usage Patterns}
To combine usage patterns for ChatGPT from \textcite{Chatterji2025} with those of Claude, we must adapt our approach, because 2025 ChatGPT usage is publicly available only at the level of 41 GWAs and not at the level of 17,659 tasks. This is a considerable limitation. \textcite{Massenkoff2026} find that occupational exposure measures are somewhat sensitive to the level of task aggregation, and our own exploratory analyses of Anthropic data aggregation to the task versus GWA level corroborate this sensitivity. GWA-level data are less precise than task-level data in that they combine many different types of tasks under a broad umbrella, but they have the advantage of being attached to GWA importance scores and level-of-sophistication scores for each job in O*NET. Still, to adjust for their reduced precision, we set GWA-level AI exposure estimates lower than our task-level exposure estimates. As shown in Table \ref{tabdec}, we assume that only 30\% of tasks in the highest decile (10) of ChatGPT work usage can be automated with modest extensions of current generative AI capabilities, followed by 25\% in decile 9, 20\% in decile 8, 15\% in decile 7, 10\% in deciles 5 and 6, and 5\% in decile 1. The latter decile includes the 18 GWAs with very low ChatGPT usage. When we average GWA exposure estimate to the occupational level, we multiply each GWA-specific AI exposure estimates by the GWA's relative importance to the job, where the importance weights across all GWAs in a job are scaled to sum to 100. The occupation-level exposure is then the occupational mean of the weighted exposure estimates across the 41 GWAs for that job. In contrast, we construct job-level exposure from the Anthropic Claude data by taking the occupational average estimated exposure across all 17,659 equally weighted tasks.

\begin{table}
	\centering
	\caption{ChatGPT query deciles and projected AI exposure}
	\label{tabdec}
	\begin{tabular}{lccc}
		\hline    
		Decile &  n GWAs  & \% of Queries & Projected Exposure \\
		\hline
		10 & 4 & 14.0 & 30\%  \\
		9 & 4 & 8.4 & 25\% \\
		8 & 4 & 2.3 & 20\%  \\
		7 & 4 & 0.9 & 15\% \\
		6 & 4 & 0.6 & 10\% \\
		5 & 4 & 0.4 & 10\% \\
		1 & 18 & 0.2 0 & 5\%  \\
		\hline 
	\end{tabular}
\end{table}

\begin{figure}[htbp]
	\centering
	\caption{Occupational automation exposure scores from five approaches}
	\label{occ_hist5u}
	\includegraphics[width=0.95\textwidth]{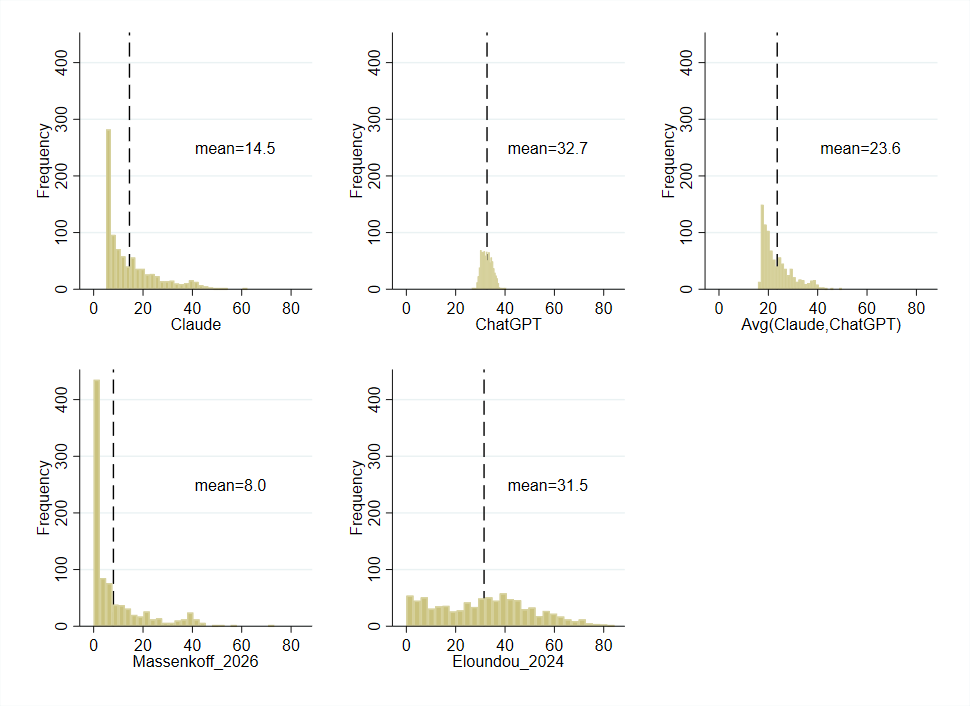}
\end{figure}

For Anthropic Claude data alone, this approach yields an occupation-level job exposure range from 5\% to 62\% with a mean of 14.5\% and a standard deviation of 10.8 percentage points. For the OpenAI ChatGPT data alone, it yields occupation-level exposure estimates ranging more narrowly from 27\% to 40\% with a mean of 32.7\% and a standard deviation of only 2.2 percentage points. These two job-level distributions are shown in the top-left and top-center panels, respectively, of Figure \ref{occ_hist5u}. Not surprisingly, the Anthropic-only exposure measure we construct from September-released data is quite similar to the measure by \textcite{Massenkoff2026}, which used the November-released data, as shown in the bottom left of Figure \ref{occ_hist5u}. Both have a sharp left peak and a strong positive skew. Our metric has a higher mean and slightly softer skew because it represents the plausible near-term exposure of occupational tasks based on recent usage rather than the percent of recently exposed tasks, as in \textcite{Massenkoff2026}. In contrast to both Anthropic-based measures, the ChatGPT exposure estimates are bell-shaped with the aforementioned smaller range.

To construct our own \textit{occupational average exposure measure} from both sources, we take a simple average of our Claude and ChatGPT exposure scores at the occupational level, with the resulting exposure distribution shown in the top-right panel of Figure \ref{occ_hist5u}. The combined measure has a mean occupational exposure of 23.6\%, a range from 15.8 to 29.8, and a standard deviation of 6.18 percentage points, as shown at the top of Table \ref{tabcomp}. This places its mean exposure below the job-level automation scores of 31.5 in \textcite{Eloundou2024}, as shown in the bottom row of Figure \ref{occ_hist5u}, but well above the present-day automation score mean of 8.0 as found in \textcite{Massenkoff2026}. In other words, it harmonizes the signal provided by the distinct data sources.

\begin{figure}
	\centering
	\caption{Standardized exposure in seven key models}
	\label{stdhist}
	\includegraphics[width=.95\textwidth]{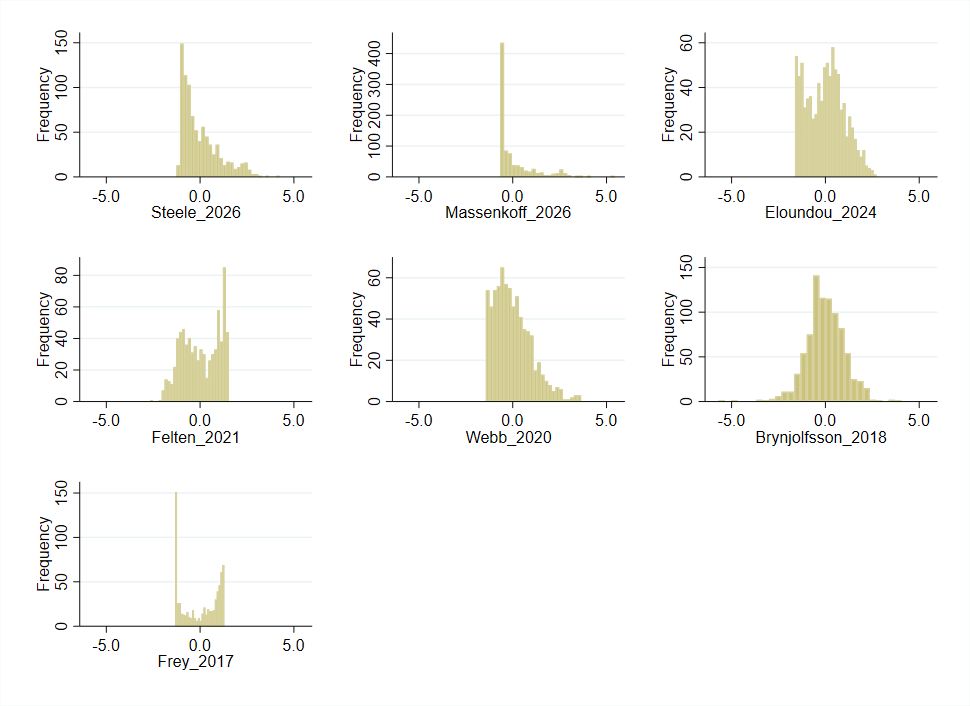}
\end{figure}

To facilitate broader comparisons among the different automation exposure estimates in this paper, we next standardize all estimates to have a mean 0 and standard deviation of 1, including our combined Claude/ChatGPT estimates, the \textcite{Massenkoff2026} estimates, and the \textcite{Eloundou2024} estimates, as well as estimates in the other models summarized above. Figure \ref{stdhist} displays the univariate distributions for all seven models. The distributions are shown with a common x-axis, but the y-axes are allowed to vary. Comparing the standardized distributions shows that they have very different shapes, but the real question of interest is whether the different estimation approaches are ranking jobs similarly in terms of their likely automation exposure. 

\subsection{Correspondence Among Occupational AI Exposure Projections}
\begin{figure}
	\centering
	\caption{Scatterplot matrix of standardized occupational exposure estimates}
	\label{occ_mat7sdu}
	\includegraphics[width=0.95\textwidth]{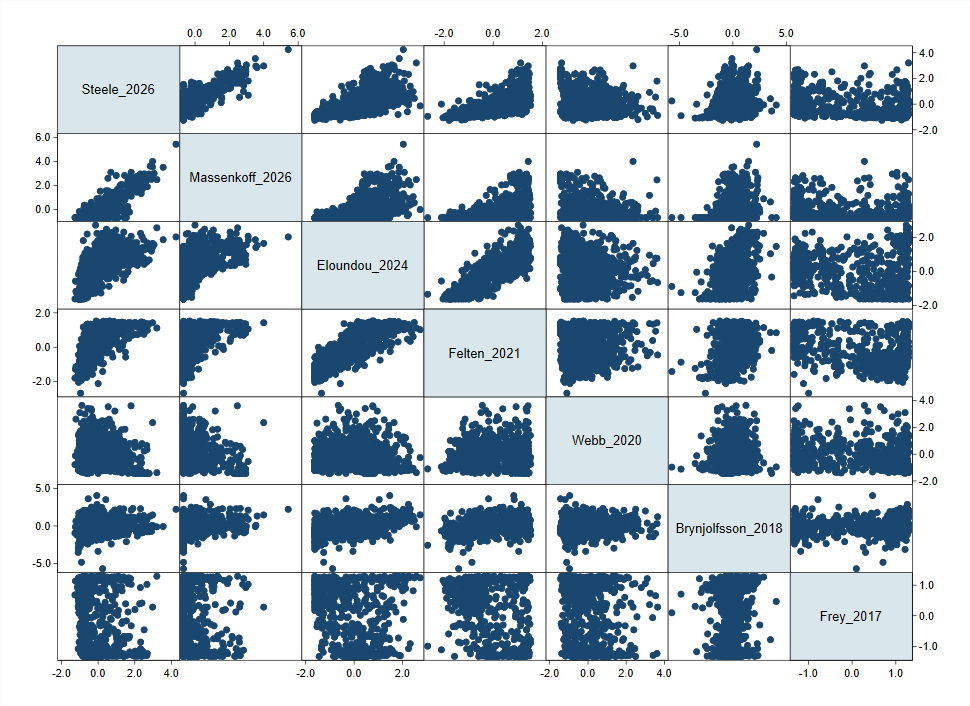}
\end{figure}

To consider this question, we turn to Figure \ref{occ_mat7sdu}. This matrix displays two-way scatterplots for each pair of models we compare in this paper, with all estimates standardized to the same scale. If the models offer similar signals of jobs' automation exposure under AI, we would expect positive, linear correlations across the matrix. But this is not what we find. First, the models do not rank jobs similarly in their AI exposure. The strongest positive correlations are found between our estimates and those of \textcite{Massenkoff2026}, because both are drawing on the empirical usage data from Anthropic, and between \textcite{Eloundou2024} and \textcite{Felten2021}, which are based on ChatGPT guesses and human projections, respectively, of tasks that generative AI  can feasibly do. The correspondence between is Eloundou et al.'s and Felten et al.'s estimates is interesting because the two models differ in their level of analysis (19,000 O*NET tasks versus 52 job abilities), their respondents (ChatGPT as rater versus Mechanical Turk human survey respondents), and their questions (fraction of the endeavor LLMs could do versus whether generative AI could do the endeavor). On the other hand, they share a focus on the theoretical capabilities of generative AI. This is distinct from focusing on what LLMs are currently being asked to do \parencite{Steele2026, Massenkoff2026}; which AI applications are being patented \parencite{Webb2020}; which tasks machine learning can do \parencite{Brynjolfsson2018}; and which jobs digital automation can do \parencite{Frey2017}. Webb's patent-based estimates, Brynjolfsson et al.'s machine-learning-based estimates, and Frey's estimates based on known automation bottlenecks show very little correspondence with each other or with later measures that considered LLM capability or usage. In general, the newer models from \textcite{Felten2021} onward show slightly higher correspondence than the older models, perhaps because the newer approaches are more sensitive to the capabilities offered by generative AI.

It is important to note that our combined Claude/ChatGPT occupational AI exposure estimates and those of \textcite{Massenkoff2026} have a correlation of 0.89. Because of this high correlation and because half of our own estimates are based on Anthropic usage data from the same year as Massenkoff's data, we include our estimates in lieu of Massenkoff's in the analyses that follow. This reduces the chance that the cross-model results are overly influenced by recent Anthropic usage. Also, we derive empirical feasibility projections from task usage deciles rather than from feasibility projections in \textcite{Eloundou2024}, making our estimates independent of the  \textcite{Eloundou2024} approach that \textcite{Massenkoff2026} uses to define feasibility.

\section{Results: Occupational AI Exposure Across Models}

\subsection{Models Vary in AI Exposure by Salary}
\begin{figure}[htbp]
	\centering
	\caption{Binned scatterplots of median salary by standardized AI exposure}
	\label{sal_scatter6sdu}
	\includegraphics[width=0.95\textwidth]{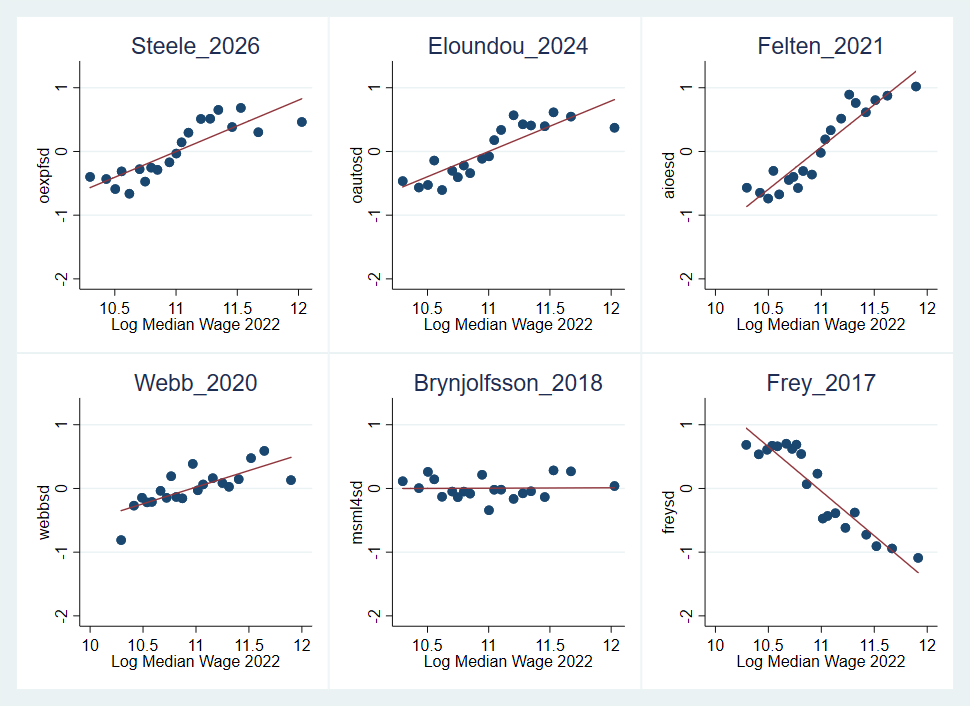}
\end{figure}

It is widely thought that AI exposure will fall hardest on white-collar jobs \textcite{Tyrangeil2026}, especially at the entry level \parencite{Beane2024}, but the extent to which the AI exposure models predict this pattern varies by the model in question. Figure \ref{sal_scatter6sdu} presents binned, standardized estimates of AI exposure against 2022 median salaries for each of the six models under consideration. The latest four models clearly show a positive, linear relationship between AI exposure and median salaries, with the survey-based ratings of \textcite{Felten2021} showing the strongest relationship with salaries. This is not surprising, since Felten et al.'s approach ascribes high automation exposure to verbal and explanatory abilities, making attorneys and educators among the most-exposed professions. The models that diverge from this pattern are \textcite{Brynjolfsson2018}, which focuses on mapping types of machine learning to tasks, and \textcite{Frey2017}, which does not use tasks, GWAs, or abilities, but focuses on jobs' probabilities of being automated based on creative, social, and dexterity bottlenecks. By partially extrapolating from prior automation waves, the estimates of \textcite{Frey2017} show a strong negative relationship with median salaries, whereas the estimates of \textcite{Brynjolfsson2018} show no relationship to salaries.

\subsection{Models Vary in AI Exposure by Occupational Level of Complexity}
Next, we instead examine the relationship between job complexity and AI exposure across models. Here, we classify job complexity not by its O*NET Job Zone but instead by rating the importance-weighted sophistication or difficulty level of each of the 41 GWAs comprising each job. This is possible because O*NET provides a rating of not only the importance of each GWA to the job (rated on a 1 to 5 scale) but also the level of sophistication at which that GWA is carried out, scored on a 0 to 7 scale. We rescale these levels of complexity to a 0-100 scale by dividing the raw level (scored by incumbents in the job) by 7 and multiplying the quotient by 100. We exclude any GWA listed as not relevant to the job, but because this pertains to only a few occupation-GWA pairings, the results are not sensitive to this choice. When weighting the GWA levels by their relative importance to the job, we divide the level by 100, placing it on a 0 to 1 scale, and multiply by the relative importance weight of 0-100. We ultimately classify a job's aggregate level of GWA complexity by aggregating its importance-weighted levels across all of its GWAs. This yields an occupational complexity variable ranging from 18.6 to 74.5, with a mean of 52 and a standard deviation of 9.

\begin{figure}[htbp]
	\centering
	\caption{Binned scatterplots of occupational complexity by standardized AI exposure}
	\label{lev_scatter6sdu}
	\includegraphics[width=0.95\textwidth]{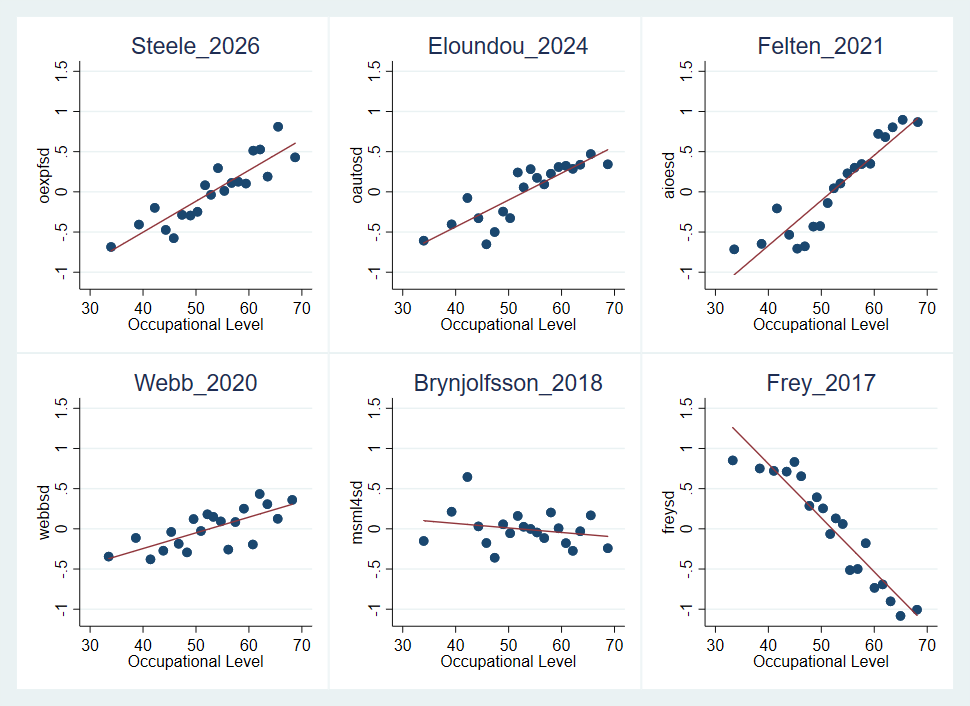}
\end{figure}

In Figure \ref{lev_scatter6sdu}, we show the binned scatterplots of AI exposure estimates against occupational level of complexity as just defined. Patterns are similar to those in the salary graph, but with occupational complexity, the differences in slopes are slightly more pronounced. AI exposure estimates from \textcite{Felten2021} show the strongest relationship to occupational complexity. The positive slope for \textcite{Webb2020} is modest, as Webb's model relies on the relationship between AI patents and O*NET job descriptions, not on the ease of automating particular tasks or skills, \textit{per se}. The relationship between AI exposure and occupational level in \textcite{Brynjolfsson2018} is slightly negative, perhaps because the machine learning rubric that raters used in \textcite{Brynjolfsson2018} rated tasks as less amenable to machine learning if they carried very high stakes. It would therefore be unsurprising if jobs with higher levels of complexity received slightly lower AI exposure scores on average. 

The takeaway of Figures \ref{sal_scatter6sdu} and \ref{lev_scatter6sdu} is that projections about the economic distributional effects of AI depend enormously on the assumptions of projection model. Relatively recent models yield very different projections from slightly older models. It is useful to understand this variation in order to consider the advantages and vulnerabilities of different prediction assumptions.

Table~\ref{topjobs} lists the top-ranked occupations for each exposure model. These differences reflect the underlying measurement philosophy of each approach: models that emphasize generative AI capabilities (e.g., Steele, Eloundou, Felten) highlight technical and analytical roles, while Webb's patent-based model highlights engineering jobs, and Frey's model emphasizing conventional automation bottlenecks highlights work that is very formulaic, including telemarketers reading sales scripts and insurance underwriters pricing risks from big datasets. Interestingly, although \textcite{Eloundou2024} and \textcite{Frey2017} use very different approaches and yield uncorrelated projections in Figure \ref{occ_mat7sdu}, they both rank telemarketers among the two professions with the highest exposure to AI.

\begin{table}[htbp]
	\centering
	\caption{Most-exposed occupations by model}
	\label{topjobs}
	\begin{tabular}{ m{0.22\linewidth} m{0.36\linewidth} m{0.36\linewidth} }
		\hline
		\textbf{Model} & \textbf{Top Occupation 1} & \textbf{Top Occupation 2} \\
		\hline
		Steele 2026 & Database Warehouse Specialist & Business Intelligence Analyst \\
		Eloundou 2024 & Telephone Operator & Telemarketer \\
		Felten 2021 & Genetic Counselor & Financial Examiner \\
		Webb 2020 & Wastewater Treatment Operator & Civil Engineer Technician \\
		Brynjolfsson 2018 & Mechanical Drafter & Mortician \\
		Frey 2017 & Telemarketer & Insurance Underwriter \\
		\hline
	\end{tabular}
\end{table}

\subsection{RIASEC Interest Category Exposure Differs Across Models}
Next, we consider how the six models differently predict the AI exposure of different occupational categories of interest. We define occupational interest categories using the RIASEC framework developed by \textcite{Holland1959}, which the O*NET Interest Profiler \parencite{Labor2026} uses as the basis of its career questionnaire and recommendations. RIASEC is an acronym for the six occupational interest categories in the framework: \textit{Realistic} jobs are practical and hands-on, including physical, mechanical, and outdoor work. \textit{Investigative} jobs are analytical, scientific, logical, and precision-oriented. \textit{Artistic} jobs are creative and expressive, using images, words, music, food, or other objects. \textit{Social} jobs focus on helping, understanding, supporting, and engaging with people. \textit{Enterprising} jobs involve business, strategy, leadership, negotiation, and competition. \textit{Conventional} jobs use systems and standards to manage information, data, and materials.

\begin{figure}
	\centering
	\caption{Number of U.S. workers by Job Zone and RIASEC Category in Millions (2022)}
	\label{riasec_empu}
	\includegraphics[width=.95\linewidth]{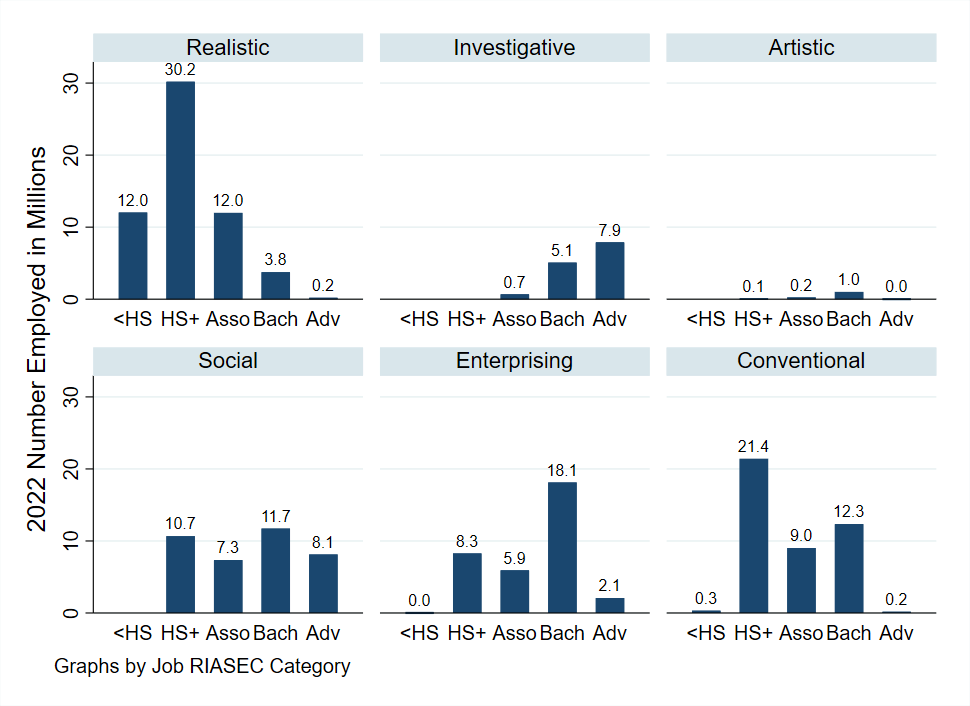}
\end{figure}

The Department of Labor scans O*NET job descriptions and rates their alignment with the six RIASEC job categories using a machine learning algorithm trained on and validated by the ratings of subject-matter experts \parencite{Putka2023}. We assign the RIASEC category with the highest rating for a particular job as the primary RIASEC category for that job. For context, Figure \ref{riasec_empu} illustrates the number of U.S. workers employed in jobs assigned to each RIASEC category in 2022 using the 749 six-digit Standard Occupational Codes (SOCs) for which workforce size is reported. We disaggregate these 189 million workers by O*NET Job Zone, describing Job Zone categories by the typical education level expected in each zone. In Figure \ref{riasec_empu} we see that a plurality of workers were employed in Realistic (manual or physical) jobs, followed by Conventional (routine office) jobs. The next most-common were Social (helping) and Entrepreneurial (business or management) jobs, in that order. Jobs in Investigative (scientific or analytical) and Artistic (creative) jobs were the least common, with Artistic jobs (e.g., film directors, graphic and video game designers, authors) accounting for only 1.3 million workers. The largest RIASEC category among bachelor's degree holders (that is, Job Zone 4) consisted of Enterprising jobs.

\begin{figure}
	\centering
	\caption{AI exposure estimates by RIASEC category for each of the six focal models}
	\label{sixmodu}
	\includegraphics[width=0.95\linewidth]{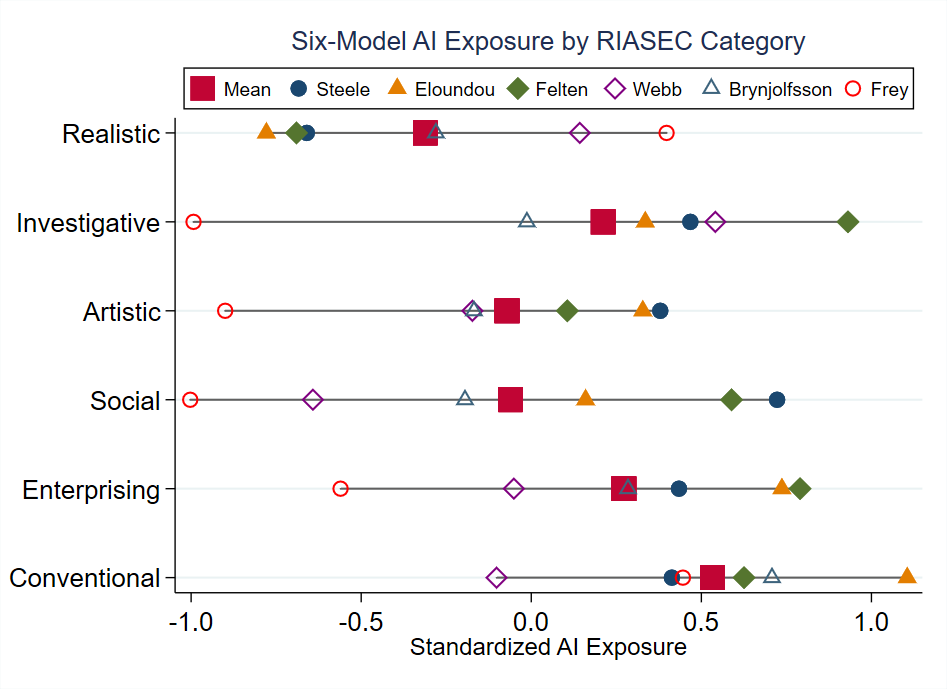}
\end{figure}

Figure \ref{sixmodu} shows automation exposure estimates from each of the six AI exposure models under consideration for each of the six RIASEC categories. The graph displays how each AI exposure model quantifies the exposure of each RIASEC category. The oldest model, \textcite{Frey2017} rate Realistic and Conventional jobs as having the highest exposure, similar to automation exposure patterns from the 1980s through 2000s \parencite{Autor2013, Autor2003}, though their model is built on an awareness of AI innovation in the 2010s. Meanwhile, \textcite{Felten2021} rates Investigative and Enterprising jobs as most highly exposed given their prevalence of communication and analytic skills. The automation scores in \textcite{Eloundou2024} rate Conventional, Enterprising, and Artistic skills as highly exposed based on ChatGPT's understanding of skills that LLMs can do independently. Our own estimates based on Claude and ChatGPT usage closely track the projections from \textcite{Eloundou2024}, except that Conventional jobs do not account for a large number of tasks automated in Claude or ChatGPT. Also, our method finds that Social jobs have the highest exposure of any RIASEC category. This rating for Social jobs would seem counterintuitive were it not for the fact that some of the most common Claude queries focused on creating lesson plans for K-12 and postsecondary teaching and explaining concepts for coaching and advising \parencite{Appel2025}. Recall also from Figure \ref{ele_clagpt2} that helping tasks in Claude were more generally augmented, meaning undertaken for validation, learning, and iteration, than fully delegated to Claude. Meanwhile, the models from \textcite{Brynjolfsson2018} provide near-average exposure across RIASEC categories. The estimates from \textcite{Webb2020}, based on patent applications' alignment with job descriptions, are anomalous in that they show high AI exposure among Realistic (physical) and Investigative jobs, and relatively low exposure in other areas. This may be due to the fact that many LLM plug-ins and wrappers are difficult to patent, whereas inventors of robotic and scientific tools are more suitable for patent applications.

The large red square in each RIASEC category represents the cross-model average of the six estimates for that category. Averaging across models provides a cross-model signal of AI exposure across the diverse sets of assumptions in each model. The cross-model averages show highest exposure in Conventional jobs, lowest exposure in Realistic jobs, and moderate exposure in Investigative and Entrepreneurial jobs.

\subsection{Cross-Model Average Robustness to Older Model Exclusion}
\begin{figure}[htbp]
	\centering
	\begin{minipage}{0.49\textwidth}
		\centering
		\caption{AI exposure estimates by RIASEC category for five most recent models}
		\label{fivemodu}
		\includegraphics[width=\textwidth]{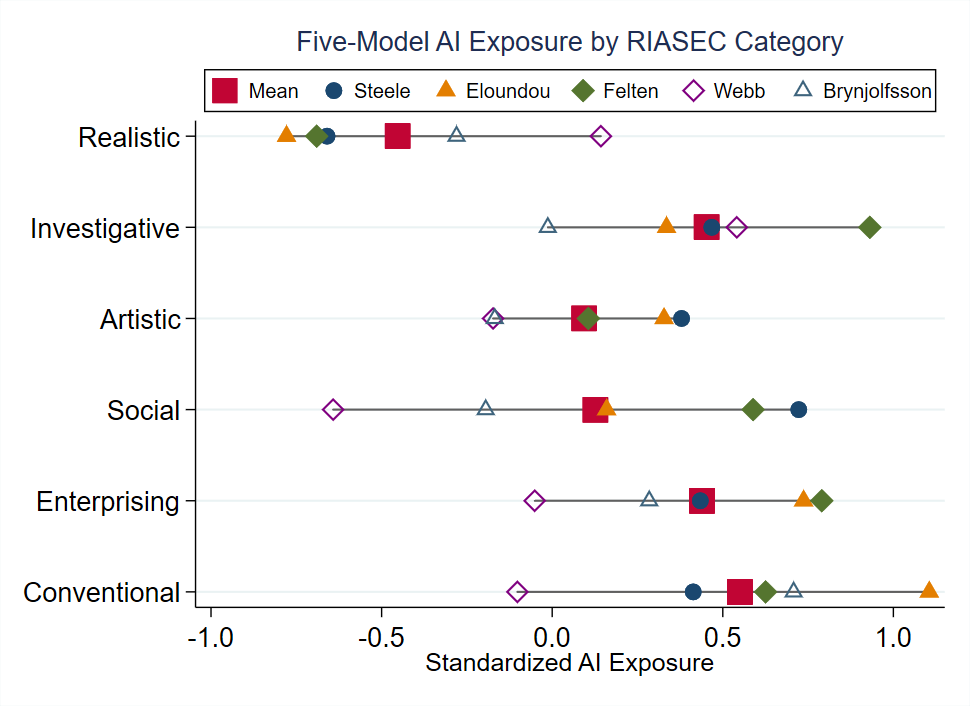}
	\end{minipage}\hfill
	\begin{minipage}{0.49\textwidth}
		\centering
		\caption{AI exposure estimates by RIASEC category for four most recent models}
		\label{fourmodu}
		\includegraphics[width=\textwidth]{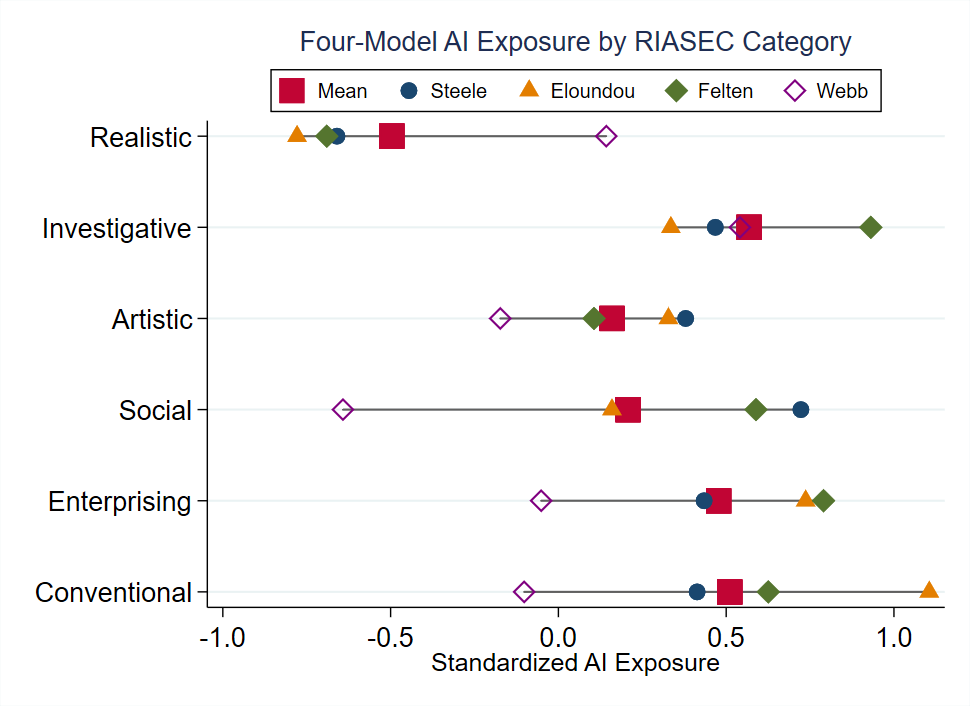}
	\end{minipage}
\end{figure}

In Figure \ref{fivemodu}, we show that the cross-model means do not change dramatically if we omit the estimates from \textcite{Frey2017}, which are the oldest and most anomalous of those we consider. The means become slightly more divergent between Realistic and the other interest categories if we omit both \textcite{Frey2017} and \textcite{Brynjolfsson2018}, as shown in Figure \ref{fourmodu}. 

Our aim in averaging across models is to provide a reliable signal that is not overly dependent on the assumptions of any particular prediction model. In our analyses that follow, which illustrate potential trade-offs between salaries and AI exposure, we elect to focus on a five-model average that excludes the anomalous \textcite{Frey2017} estimates, consistent with estimates in Figure \ref{fivemodu}. It is important to note, however, that including all six models, with \textcite{Frey2017} included, yields quite similar exposure recommendations overall.

\subsection{Balancing Trade-Offs Between Salaries and AI Exposure}
Given that salaries and occupational sophistication are positively correlated with AI exposure across four of the six prediction models, we next provide guidelines for helping career seekers and career coaches, including high school students and their counselors, to conceptualize likely trade-offs between salaries and AI exposure. To do so, we rank occupations according to their median salaries in 2022, extrapolating to all 863 eight-digit occupations in our analysis based on the salaries of their six-digit Standardized Occupational Codes. We also rank occupations by their standardized AI exposure scores averaged across the five most-recent models considered in Figure \ref{sixmodu}, meaning all \textit{except} the estimates from \textcite{Frey2017}. Based on these separate job rankings on salary and AI exposure, we classify occupations into four categories:

\begin{enumerate}
	\item \textit{High salary, Low AI exposure} are jobs that pay above the median salary and also score at or below the median of the cross-model average AI exposure scores. This category is poised for relative security and stability in the AI-enabled economy.
	\item \textit{High-salary, High AI exposure} are jobs that pay above the median salary but also have above-median AI exposure. This category may appeal to flexible individuals seeking career dynamism in higher-paying fields, even if the mix of tasks is subject to change.
	\item \textit{Low-salary, Low AI exposure} are jobs that score at or below the median on both salaries and AI exposure. This relatively large category may appeal to individuals willing to sacrifice higher salaries for greater stability in their fields.
	\item \textit{Low-salary, High AI exposure} are jobs that pay at or below the median and have above-median AI exposure. This category is likely the most vulnerable in the AI-enabled economy, though it may offer stepping stones for career changers seeking to grow, learn, and adapt.
\end{enumerate}

\begin{figure}
	\centering
	\caption{Salary versus five-model AI exposure by RIASEC category}
	\label{exposure_payu}
	\includegraphics[width=0.98\linewidth]{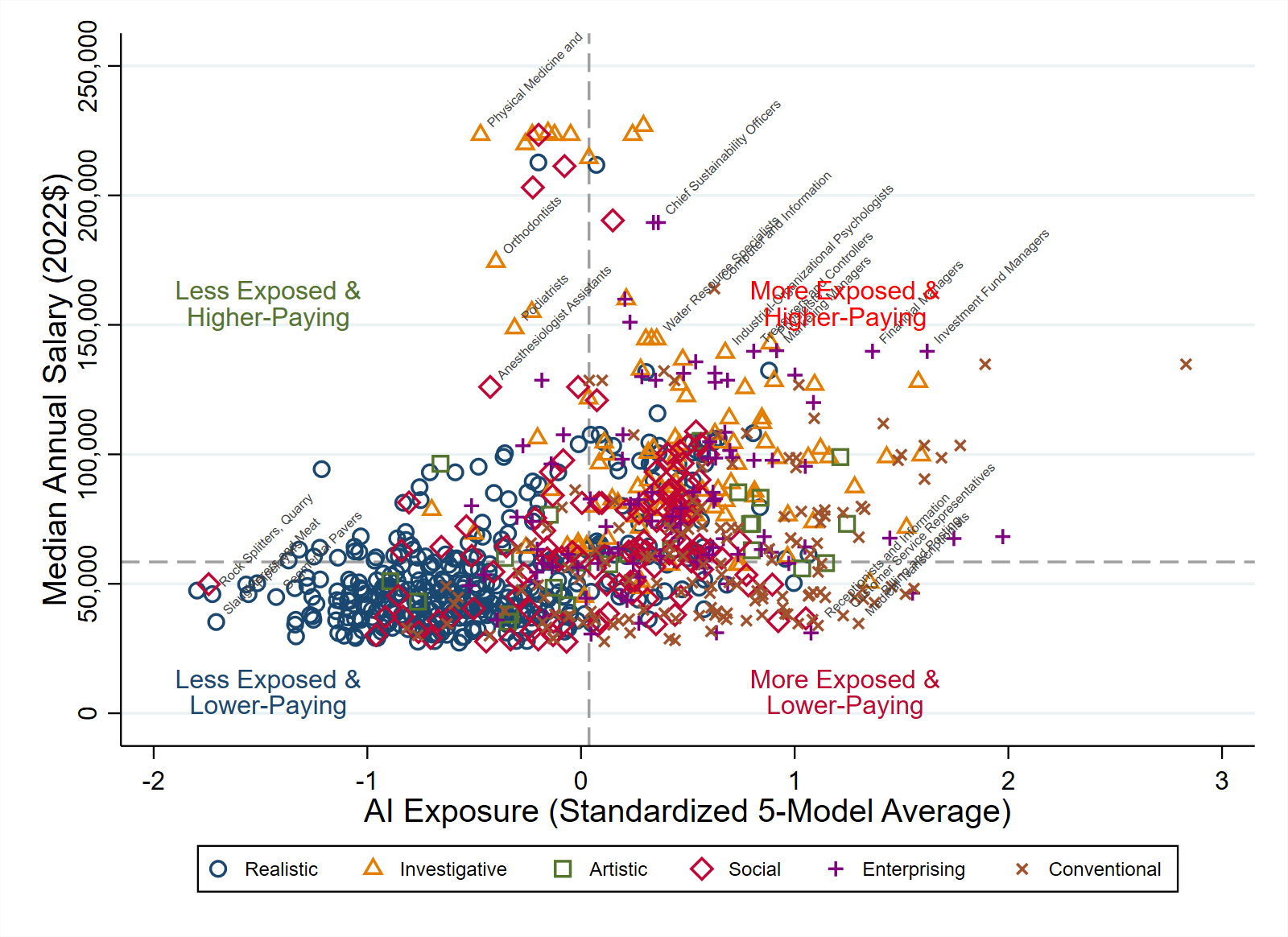}
\end{figure}

In Figure \ref{exposure_payu}, we display a scatterplot of all 863 jobs with salary data, ranked according to their standardized cross-model AI exposure on the x-axis and their median salaries on the y-axis. The shapes and colors of the markers indicate the primary RIASEC category of each job. A few jobs are labeled to provide context about the types of occupations in each quadrant. The top-left quadrant represents jobs with high salaries and low AI exposure. The top-right quadrant represents high salaries and high AI exposure, and the bottom two quadrants represent jobs paying below-median salaries, with AI exposure increasing from left to right. What is clear from the graph is that median salaries rise modestly with AI exposure, on average, consistent with our having averaged the AI exposure projections from the first five models in Figure \ref{sal_scatter6sdu}. Also, below-median salaries have a compressed range from about \$28K to \$58K in 2022 dollars, whereas above-median salaries represent a broader salary range. The highest-paying jobs, which tend to be medical doctor or chief executive roles, fall near the median in terms of projected AI exposure. (Readers wishing to examine the scatterplot in greater detail and compare individual exposure models to the cross-model average can do so with our interactive data tool available at \url{https://tinyurl.com/SteeleCruzCareersAI}.)

\begin{figure}
	\centering
	\caption{Salary and AI exposure intersection by RIASEC category}
	\label{barstack_riahiu}
	\includegraphics[width=0.95\linewidth]{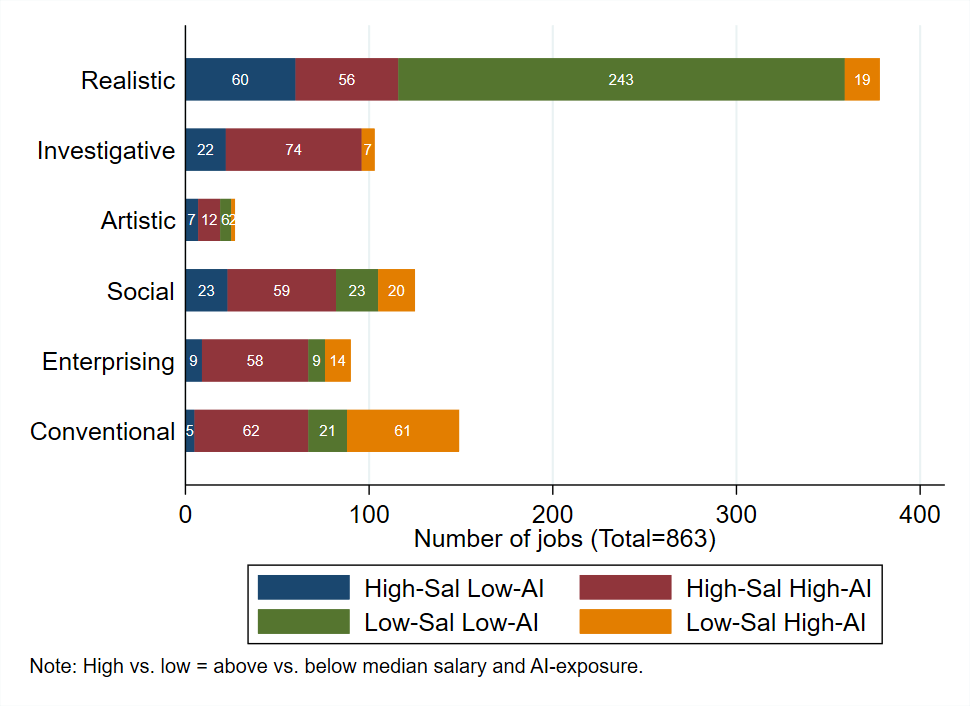}
\end{figure}

In Figure \ref{barstack_riahiu}, we show the number of jobs in each salary-by-AI-exposure category, disaggregated by RIASEC category. The blue category in the graph is considered optimal for safety and stability in that it matches higher-paying jobs with relatively low  AI exposure predictions. The green category is considered stable but less remunerative, as it pairs low projected AI exposure with below-median pay. For workers eager to pivot or adapt their work over time, the red category pairs higher-than-median pay with higher-than-median projected AI exposure. The Realistic category (physical and manual work) accounts for the largest number of occupations, more than half of which are classified as having low exposure to AI. Jobs that are projected to have low AI exposure along with above-median salaries are found primarily in the Realistic, Investigative, and Social categories.

\begin{figure}
	\centering
	\caption{Salary and AI exposure intersection by O*NET Job Zone}
	\label{barstack_jobzoneu}
	\includegraphics[width=0.95\linewidth]{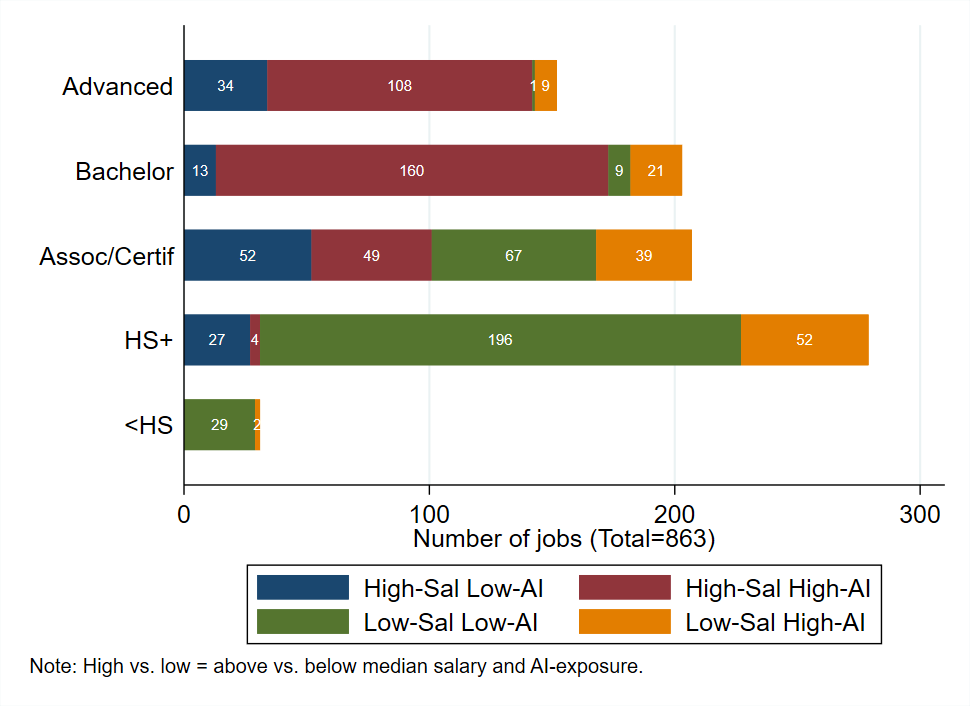}
\end{figure}

In Figure \ref{barstack_jobzoneu} we consider how salaries and AI exposure vary by O*NET Job Zone. Job Zones are not strictly measures of educational requirements but of the combined level of education, experience, and on-the-job training expected for a given occupation \parencite{Labor2026b}. In practice, however, Job Zones clearly differ in average education levels, so Figure \ref{barstack_jobzoneu} uses education level as categorical shorthand for occupational skill level. Job Zone 1, for instance, is free of education and experience requirements; these jobs may be staffed by people with less than a high school diploma. The highest Job Zone, 5, requires elevated skills and experience consistent with an advanced degree. Figure \ref{barstack_jobzoneu} suggests that the Job Zone with the largest share of high-paying, low-AI exposure jobs is Zone 3. This corresponds to associate's degree holders or skilled laborers without bachelor's degrees, such as plumbers, carpenters, dental hygienists, or phlebotemists. The cross-model average AI exposure appears to be highest at the bachelor's degree level. Jobs with low AI exposure and below-median salaries are concentrated in the three lowest Job Zones.

\begin{figure}
	\centering
	\caption{Salary and AI exposure intersection by occupational field}
	\label{barstack_socmaju}
	\includegraphics[width=0.98\linewidth]{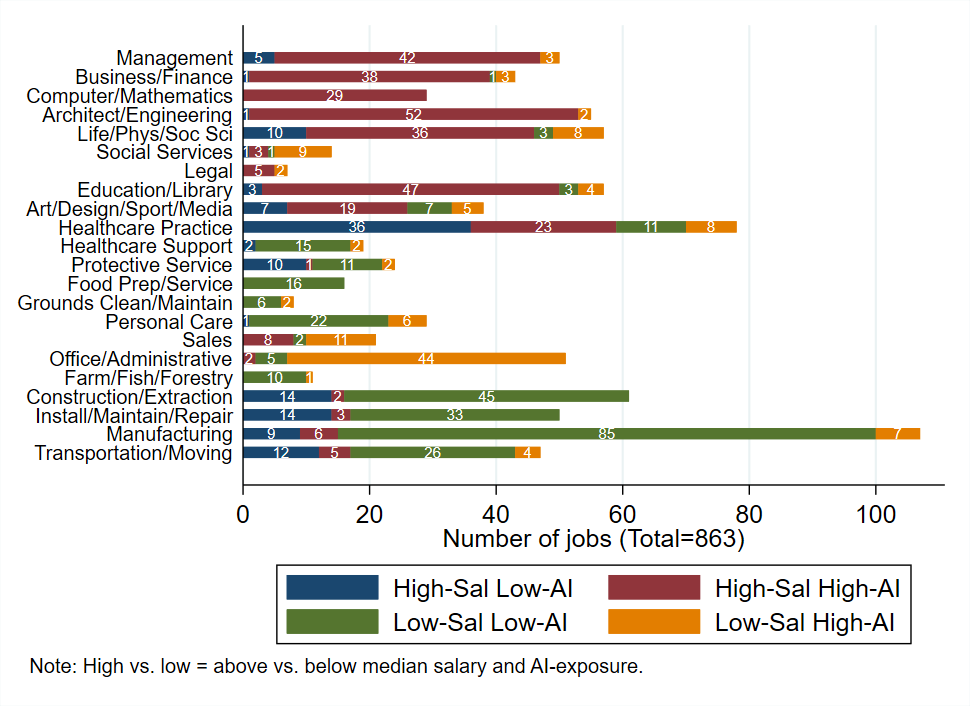}
\end{figure}

In Figure \ref{barstack_socmaju}, we display the salary and AI exposure intersections across 22 occupational fields, defined by the first two digits of their Standard Occupational Codes. The field with the largest number of jobs in the high-paying, low-AI exposure category is healthcare practice, which includes medical doctors, nurses, pharmacists, veterinarians, dietitians, speech/language pathologists, sonographers, and various types of physical therapists and psychotherapists. Healthcare support roles, which consist mainly of medical assistants and nursing aides, are rated as having low AI exposure but also below-median salaries. Fields that have been thought of as relatively reliable pathways in recent decades, including management, finance, computing, engineering, law, and education are classified as paying above median salaries but having higher-than-median projected AI exposure. If these high-exposure fields do experience changes in demand or task composition, workers may have to adapt or become entrepreneurial, as many journalists (now Substackers and podcasters) learned to do in the wake of the internet revolution \parencite{Bhuller2023}. The field of office and administrative work, though lower-paying, also appears to be highly exposed to AI.

\subsection{Salaries by Augmented Versus Automated Claude Use}
\begin{figure}
	\centering
	\caption{Mean salary by augmented versus automated Claude usage decile}
	\label{augautdec}
	\includegraphics[width=0.9\linewidth]{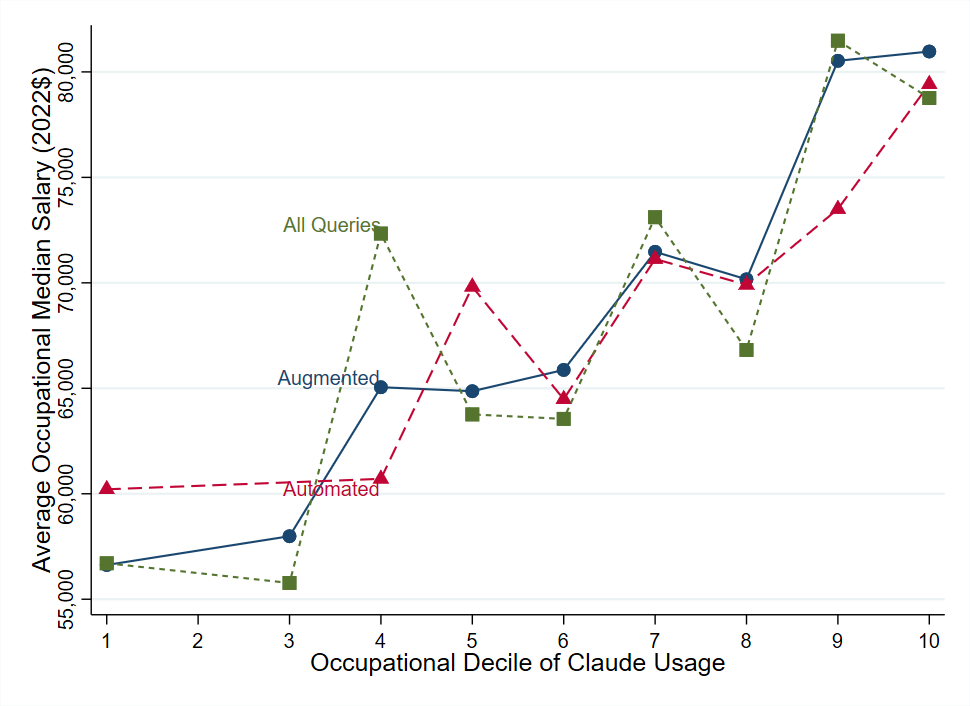}
\end{figure}

Finally, we consider whether higher-paying jobs are using generative AI in a different way than lower-paying jobs. To address this question, we return to the Claude usage data and aggregate the number of augmented queries versus automated queries to the level of the occupation. We rank 872 occupations separately on each usage type and then group the jobs into deciles in terms of their percentage of augmented (human-in-the-loop) queries and automated (task-delegation) Claude queries relative to all queries. In Figure \ref{augautdec}, we show the average salary of each augmentation, automation, and combined query decile at the occupational level. Jobs with augmentation at the ninth and tenth deciles, denoted by circles with solid-line connectors, were higher paying than those in the same decile of automation, denoted by triangles with long-dash connectors. This difference reaches \$7,000 in the ninth decile, where Claude queries are common. In other words, jobs in which Claude was used frequently as a helper in August 2025 were higher-paying, on average, than those in which it was used frequently as a delegate. This pattern could suggest that the highest-paid work remains difficult to delegate, but it should not be overemphasized for two reasons. First, in the September 2025 Anthropic data release, it is not possible to distinguish work use from personal use, so task usage is simply assigned to jobs that draw on those tasks. Second, our salary data are from 2022 so would not capture any compensation changes attributable to AI. Even so, the question how AI is being deployed in jobs with higher versus lower pay and complexity remains an important economic question.

\subsection{Discussion and Implications}
By describing and comparing seven conceptually distinct AI exposure predictions, including the set of prediction we construct from Anthropic and OpenAI usage data, we show that occupational AI exposure projections differ markedly due to variation in researchers' assumptions and methods. Models released from 2021 onward, even a year before the November 2022 release of ChatGPT 3.5, show greater consistency of projections than older models, likely because they draw on empirical or theoretical understanding of generative AI. Still, all of the models shed light on the changing demarcation line between work that machines can and cannot do.

Unlike prior waves of automation documented by \textcite{Autor2003} and \textcite{Autor2013}, AI-exposure models released since 2020 show a positive relationship between AI exposure and salaries, as well as positive relationships between AI exposure and occupational complexity as measured by the importance-weighted level of Generalized Work Activities in the job. The two older sets of projections we explore in this paper---those by \textcite{Frey2017} and \textcite{Brynjolfsson2018}---show negative or neutral relationships to salaries and occupational complexity. This may reflect their assumptions about tasks that could be automated and, in the case of \textcite{Brynjolfsson2018}, the assumption that very high-stakes takes are not suitable for machine learning. More recent models have not made assumptions about the stakes or urgency of tasks in projecting AI exposure, and society's insistence on human oversight for very high-stakes decision-making remains to be seen. The patent-text-mining approach by \textcite{Webb2020} projects higher exposure for occupations in engineering and applied sciences. This projection is consistent with his focus on the types of AI usage that are suitable for patents---often related to hardware innovation---relative to the process-oriented innovations facilitated by chatbots and their agentic extensions.

Models published from 2021 onward, including our own empirical model introduced in this paper, show lower AI exposure among Realistic (i.e., physical) occupations, and higher exposure among Investigative, Artistic, and Entrepreneurial jobs. Anthropic usage data, in particular, point to high exposure among Social jobs such as teaching, coaching, and advising. On the other hand, a disproportionate share of Social usage is augmented rather than fully automated, meaning Claude is used interactively for guidance, rather than to undertake tasks on its own. This is noteworthy because jobs that used Claude frequently as a human-guided assistant in August 2025 had higher salaries than those that used it frequently for delegation. These Claude data, it should be noted, were collected after the public release of Claude Code in May 2025 but before its autonomous coding abilities notably improved in early 2026. \parencite{Mollick2026}.

Even so, the positive association between salaries and projected AI exposure since 2020 raises questions about which financially stable career paths are also comparatively insulated from AI-induced change. By averaging across  five of the most-recent AI exposure models, including our own empirical approach, we attempt to reduce the estimation uncertainty that accompanies each set of AI exposure projections on its own. Using cross-model averages, we find that occupations that pair above-median compensation with below-median AI exposure may be found most commonly in:

\begin{itemize} 
	\item Realistic occupations, followed by Social and Investigative occupations 
	\item Jobs at the associate degree level, such as skilled trades
	\item Professions related to healthcare practice, such as medicine and nursing
\end{itemize} 

Meanwhile, higher-paying fields such as finance, computing, engineering, and law show above-median AI exposure in most jobs, suggesting that these paths may require flexibility and adaptation as the composition of tasks responds to AI innovation.

In considering the implications of these AI exposure projections, is important to emphasize that high AI exposure does not necessarily mean a job will face reduced demand. As explained above, Jevons paradox suggests that employment in AI-exposed fields may increase as the price of performing the work falls, unlocking price-sensitive demand. But even when Jevons paradox applies, AI-enabled jobs may require a level of adaptability that appeals to some workers more than others. An oft-cited example of the paradox lies in the growth of bank teller jobs between 1975 and 2010, despite the concomitant rise of Automated Teller Machines (ATMs) \parencite{Bessen2019}. The explanation is that ATMs led to the proliferation of bank branches, which hired tellers not to accept deposits but, rather, to sell additional banking and investment products. The parable is true, but it overlooks not only the two-thirds decline in bank teller jobs that followed the rise of mobile banking on smartphones in the 2010s \parencite{Oks2026}, but also the fact that a teller who is expected to sell investment products on commission has a very different job from a teller who is expected to cash checks. In other words, even as teller jobs expanded in ATM-equipped branches, the job required workers to pivot their skill sets. Some workers may find themselves galvanized by the opportunity to pivot, while others may feel overwhelmed or alienated. For this reason, it is important to think of an occupation's projected AI exposure as both a risk and a potential opportunity.

Our paper attempts to help career seekers and their career counselors to prepare for the uncertainties of the AI economy by understanding occupations that are likely to be most and least exposed to AI. Averaging across five AI projection models that draw on diverse assumptions, we show that probable AI exposure intersects with salaries, occupational interest categories, education levels, and occupational fields. Even if AI exposure does not necessitate job displacement, higher AI exposure may lead to a shift in task composition and a need to adapt quickly to new ways of working. For job seekers who prioritize safety, less-exposed jobs in higher-paying fields may offer a higher probability of stability. For those who seek novelty and are open to adaptation, the relatively high-paying business, technological, and analytic fields that have thrived in recent decades remain viable pathways. Change is likely in these occupations, but the direction it takes will be shaped in part by the AI usage priorities of the generation now entering these fields.

\section*{Supplementary Materials}
\textbf{Tableau interactive data tool for career seekers:} \url{https://tinyurl.com/SteeleCruzCareersAI}

\vline

\noindent\textbf{Data repository:} \url{https://github.com/jensteele1/careers-ai-data}

\newpage

\printbibliography

\newpage

\appendix

\section{Appendix}
% before the first appendix figure
\setcounter{table}{0}
\renewcommand{\thetable}{A\arabic{table}}

\begin{table}[htbp]
	\centering
	\caption{Country-level relationship of AI optimism to labor force attributes}
	\label{tabregress}
	\begin{tabular}{lccc} \hline
		& (1) & (2) & (3) \\
		VARIABLES & \% Optimistic & \% Optimistic & \% Optimistic \\ \hline
		&  &  &  \\
		\% Postsec Degrees & -0.290*** & -0.183 & 0.0307 \\
		& (0.102) & (0.122) & (0.0984) \\
		Avg. sr. salary 100Ks &  & -9.280 & -15.36*** \\
		&  & (6.047) & (4.612) \\
		Ratio jr. professional to min. wage &  &  & 5.208*** \\
		&  &  & (0.965) \\
		Constant & 54.76*** & 55.78*** & 38.74*** \\
		& (4.712) & (4.670) & (4.680) \\
		&  &  &  \\
		Observations & 37 & 37 & 37 \\
		R-squared & 0.188 & 0.241 & 0.597 \\ \hline
		\multicolumn{4}{p{0.9\textwidth}}{Dependent variable is \% in country who believe AI benefits $>$ risks, based on Gillespie et al. (2025) survey of $\approx$ 1000 per country merged with N26 (2022) salary and education data} \\
		\multicolumn{4}{l}{ Standard errors in parentheses} \\
		\multicolumn{4}{l}{ *** p$<$0.01, ** p$<$0.05, * p$<$0.1} \\
	\end{tabular}
\end{table}
	
\end{document}